\documentclass[prd,twocolumn,showpacs,preprintnumbers,amsmath,amssymb]{revtex4}

\usepackage{mathrsfs} 

\usepackage{bm}

\usepackage{color}
\usepackage[usenames,dvipsnames,svgnames,table]{xcolor}

\newcommand{\mr}{\mathrm}
\newcommand{\ms}{\mathscr}
\newcommand{\mc}{\mathcal}
\newcommand{\f}[2]{\frac{#1}{#2}}
\newcommand{\df}[2]{\frac{\partial #1}{\partial #2}}
\newcommand{\kron}[2]{\delta^{#1}_{\phantom{#1}#2}}
\newcommand{\ie}{{\it i.e. }}
\newcommand{\eg}{{\it e.g. }}

\newcommand{\etal}{{\it et al.~}}

\begin{document}

\preprint{Preprint number: XXXX}

\title{Simple, explicitly time-dependent and regular solutions of the linearized vacuum Einstein equations on a null cone }

\author{Thomas M\"adler}
\email{thomas.maedler@obspm.fr}
 \affiliation{Laboratoire Univers et Theorie (LUTH), CNRS/Observatoire de Paris, Universit\'e Paris Diderot, \\
5 place Jules Janssen, 92195 Meudon cedex, France.}

\date{\today}

\begin{abstract}
Perturbations of the linearized vacuum Einstein equations on a null cone in the Bondi--Sachs formulation of General Relativity can be derived from a single master function with spin weight two, which is related to the Weyl scalar $\Psi_0$, and  which is determined by a simple wave equation. Utilizing a standard spin representation of the tensors on a sphere and two different approaches to solve the master equation, we are able to determine two simple and explicitly time-dependent solutions. Both solutions, of which one is asymptotically flat, comply with the regularity conditions at the vertex of the null cone.  For the asymptotically flat solution we calculate the corresponding linearized perturbations,  describing all  multipoles of spin-2 waves that propagate on a Minkowskian background spacetime. We also analyze the asymptotic behavior of this solution at null infinity using a Penrose compactification, and calculate the Weyl scalar, $\Psi_4$. Because of its simplicity, the asymptotically flat solution presented here is ideally suited for testbed calculations in the Bondi--Sachs formulation of numerical relativity. It may be considered as a sibling of the well-known Teukolsky--Rinne solutions, on spacelike hypersurfaces,  for a metric adapted to null hypersurfaces. 
\end{abstract}

\maketitle

\section{Introduction}
Exact solutions of the Einstein field equations  provide a deeper insight into the classical theory of General Relativity. In this article, we present an exact time-dependent global solution on a null cone describing all multipoles of linearized spin-2 fields propagating on a Minkowskian background space by using the Bondi--Sachs formulation \cite{Bondietal1962,Sachs1962,TW,WinicourLRR} of General Relativity. The solution  is given as a spectral series with respect to spin-2 harmonics, where the coefficients are  simple rational expressions of the time and radial coordinate. Therefore, we refer to it as SPIN-2. It is an ideal textbook solution allowing one to demonstrate, when working in the Bondi--Sachs frame work of General Relativity,   important features, such as the regularity conditions at the vertices, the commonly used $\eth-$formalism, and the subtleties at null infinity. In addition, since it describes all radiation multipoles,  SPIN-2 is also  well suited as a testbed solution for numerical relativity, when the Einstein field equations are solved in a Bondi--Sachs framework, (see \cite{WinicourLRR} for a review). Thus it might be considered as a sibling, in null coordinates, of the well-known Teukolsky-Rinne solutions  \cite{Teuk,Rinne} employed in numerical relativity using the  $(3+1)$ formulation of General Relativity.

Despite its simplicity, this is the first time that a regular and asymptotically flat solution of the linearized vacuum Einstein equation has been reported for all multipoles in Bondi--Sachs formulation. Linearized solutions were first discussed qualitatively by Bondi \etal \cite{Bondietal1962}, who gave an asymptotic vacuum solution in terms of inverse powers of an areal distance coordinate $r$  of an axisymmetric metric with a hypersurface orthogonal Killing vector. As their solution was given by coefficients of a series of $r^{-n},\,(n>0)$, it is not regular at the vertices of the null cones. Winicour \cite{W1983,W1984,W1987} proposed a  Newtonian approach to the Bondi--Sachs formulation of General Relativity which is related to a  post-Minkowskian expansion of the Bondi--Sachs metric. In particular, he pointed out  the necessity of imposing regular boundary conditions at the vertex of a freely falling Fermi observer in the background spacetime, and he introduced spin-0 potentials \cite{N2BMS1966} to solve for the perturbations.  Axisymmetric linearized solutions were revisited by Papadopoulos \cite{PapPhD,Gomez1994}, who used Winicour's idea of the spin-0 potentials to find solutions of linearized vacuum perturbations.  Papadopoulos's algorithm was later generalized  by Lehner \cite{LuisPhD,CCE1996} to three dimensions.   

The spin-0 potential approach to find a radiative $l-$multipole can be summarized in three basic steps:   First; one guesses  a regular solution at the vertex for a monopole scalar field  that obeys the flat space wave equation. Second; one applies $n-$times the  $z-$translation operator expressed in outgoing polar null coordinates \footnote{The $z-$translation operator in outgoing polar null coordinates is $\partial_z = \cos(\theta)(\partial_u-\partial_r)-r^{-1}\sin\theta \partial_\theta$.} to the monopole solution to find a $n-$multipole of the scalar wave equation. These multipoles are also a solution of the scalar wave equation, because the $z-$translation operator commutes with the axially symmetric angular momentum operator. Finally; one finds the Bondi--Sachs metric functions by applying the $\eth-$ operator \cite{N2BMS1966, eth} to the $n-$multipoles solution of the scalar wave equation and  integrates it to obtain the Bondi--Sachs metric functions.  This elegant method has the disadvantage that it  requires an infinite application of the $z-$translation operator  for generating a  solution for all $n-$multipoles. 

Linearized and quadratic perturbations, with respect to a Minkowski background, were considered by \cite{GN1997}, who presented their solutions in terms of Newman--Penrose quantities \cite{NP1962} and, like Bondi \etal \cite{Bondietal1962}, they  gave only leading order terms of a $r^{-1}$ expansion of these quantities in the asymptotic regime. Bishop and collaborators \cite{B2005, BK2009} proposed a  procedure to find linearized perturbations on a Schwarzschild background. However,  in their approach ``...some of the expressions get very complicated and in order to simplify the presentation we now specialize to the case $l=2$...'' \cite{B2005} to obtain the solution is overly complicated, although the calculation of linearized perturbations is achievable by a simpler method. Nevertheless, Reisswig \etal \cite{RPB2012} determined a $l=3$ multipole using Bishop's approach. 

We solve the vacuum Einstein equations  for the zeroth and first order terms of an expansion of the metric in terms of a measure of the deviation from spherical symmetry. Thereby we assume that the null cones emanate from a Fermi observer \cite{MM1963} following the timelike geodesic of the background spacetime. As the boundary conditions at the vertices  are given by the regularity conditions of the Fermi observer \cite{MM2013}, we determine these boundary conditions for the general three dimensional case using spin weighted harmonics. Utilizing these boundary conditions, we then integrate the Einstein equations from the vertex to infinity employing two different approaches: one in which an asymptotically flat solution is obtained whereas in the other one the solution diverges exponentially. Applying the Penrose compactification \cite{Penrose1963} to the asymptotically flat solution,  the Weyl scalar $\Psi_4$ at null infinity is calculated with the formalism of \cite{Babiuc2009}.

This article is organized as follows: In Sec. \ref{sec:MinkCone}, we introduce the quasi spherical approximation of a Bondi--Sachs metric. In Sec. \ref{sec:sol_proc}, we derive the necessary equations and framework to find solutions of the zeroth and first-order quasi-spherical approximation. In particular, we  introduce a  function, which we call {\it master function}, which is related with the Weyl scalar $\Psi_0$, and whose solution allows us to determine the linearized perturbation of a vacuum spacetime. We also present equations obtained with the $\eth-$ formalism that will be used to find a solution of the perturbations. In Sec. \ref{Sec:RegPsi}, we determine the general boundary conditions of the master function and the resulting boundary conditions for the perturbations in three dimensions using spin-spherical harmonics. In Sec. \ref{sec:2sols}, two solutions of the master function are presented, one of which diverges exponentially as $r$ tends to infinity whereas the other one is asymptotically finite. For the asymptotically finite solution of the master function, we calculate the resulting perturbations and  the Weyl scalar $\Psi_4$ at null infinity in Sec. \ref{sec:RegSol} adopting the  formalism of \cite{Babiuc2009}. Finally, our results are summarized in Sec. \ref{sec:Discuss}.

We use geometrized units in our calculations and the conventions of \cite{MTW} for curvature quantities. For tensor indices, we use the Einstein sum convention, where small Latin indices $(a,b,c,...)$ take values $0,...,\,3$ and capital Roman indices $(A,\,B,\,C,\,...)$ have values $2,\,3$ corresponding to angles $\theta$ and $\phi$, respectively. Bondi--Sachs coordinates are denoted by $x^a$, conformal Bondi--Sachs coordinates by $\hat{x}^a$,  and conformal inertial coordinates by $\tilde{x}^a$. A quantity calculated in conformal Bondi--Sachs coordinates is denoted by a hat over the respective quantity,  one in conformal coordinates  by a tilde, and a complex conjugated quantity by an over-bar. The expression $A\hat{=}B$ means that $A-B=O(\varepsilon^2)$ in a quasi-spherical approximation. 
\section{Quasi-spherical approximation of the Bondi--Sachs metric}\label{sec:MinkCone}
Consider a smooth one-parameter family of  metrics $g_{ab}(\varepsilon)$ at null cones in a vacuum spacetime, where $\varepsilon$ is a parameter that measures deviations from spherical symmetry, and when $\varepsilon=0$ we recover the metric of a spherically symmetric spacetime. This spherically symmetric spacetime is referred to as background spacetime and contains a unique geodesic, the central geodesic, which traces the centers of symmetry.  The metrics $g_{ab}(\varepsilon)$ are expressed in terms of Bondi--Sachs coordinates, $x^a$, \cite{Bondietal1962,Sachs1962,TW}. Given the central timelike geodesic of the background spacetime, the coordinate $x^0=u$ is the proper time along the geodesic and is constant along outgoing null cones for points that are not on this geodesic.  The two coordinates $x^A=(x^2,\,x^3)$ are angular coordinates parameterizing spheres centered on points on the timelike geodesic. Finally, the coordinate $x^1=r$ is an areal distance coordinate, such that surfaces $dr=0=du$ have the area $4\pi r^2$. The line element of the metric is given by
\begin{eqnarray}
\label{eq:lineelement}
ds^2 &=& -e^{2\Phi(\varepsilon)+4\beta(\varepsilon)}du^2 - 2e^{2\beta(\varepsilon)}dudr \nonumber\\
&&\!\!\!+r^2 h_{AB}(\varepsilon)[dx^A -U^A(\varepsilon)du][dx^B -U^B(\varepsilon)du]\;\;.
\end{eqnarray}
where $h^{AC} = h_{CB}=\kron{A}{B}$ and $|h_{AB}|_{,r}=0=|h_{AB}|_{,u}$ \cite{TW}. The latter condition on  the metric 2-tensor $h_{AB}$ is a consequence of the requirement of the radial coordinate to be an areal distance coordinate. This implies that $h_{AB}$ has only two independent degrees of freedom to describe the geometry of the 2-surfaces $dr=0=du$. 
The metric functions in \eqref{eq:lineelement} are assumed to obey an expansion in terms of the parameter $\varepsilon$ like
\begin{eqnarray}
\label{eq:def_perturb}
\beta &=& \beta_{(0)} +\sum^\infty_{n=1} \beta_{(n)}\varepsilon^n\;\;,\quad
\Phi = \Phi_{(0)} +\sum^\infty_{n=1} \Phi_{(n)}\varepsilon^n\;\;,\quad\nonumber\\
h_{AB} &=& q_{AB} +\sum^\infty_{n=1} \gamma^{(n)}_{AB}\varepsilon^n\;\;,\quad
U^A= \sum^\infty_{n=1}U^A_{(n)}\varepsilon^n\;\;,
\end{eqnarray}
where $q_{AB}$ is a unit sphere metric with respect to the coordinates $x^A$, and we denote its associated covariant derivative by $D_A$, \ie $D_A q_{BC} =0$.  Note  the symmetric tensors $\gamma_{AB}^{(n)}$ contain only two degrees of freedom and are traceless because of the determinant condition on $h_{AB}$.
The metric functions have to obey regularity conditions at the vertices of the outgoing null cones, because the vertices trace the origin of a Fermi normal coordinate system \cite{MM1963} along the central geodesic. According to the vertex lemma in \cite{MM2013} that lists the regularity properties of a Bondi--Sachs metric at the vertex, we require  the metric functions to have the following limiting behavior at $r=0$
\begin{subequations}
\begin{eqnarray}
O(\varepsilon^0) &\!\! :\!\! &\{ \beta_0,\Phi_0\}=O(r^2)\;\;,  \label{eq:reg_back}\\
\!\!\!\!\!\!\!O(\varepsilon^n) &\! \!:\!\! &\{ \gamma^{(n)}_{AB},\,\beta_{(n)},\Phi_{(n)}\}=O(r^2)\,,\;\,
     U^A_{(n)}=O(r) .\label{eq:reg_pert} 
\end{eqnarray}
\end{subequations}

Since the metric functions are given by a power series in terms of $\varepsilon$, any quantity derived from the metric $g_{ab}(\varepsilon)$ is given by a power series in terms of $\varepsilon$. For any tensor  $\mc{T}(\varepsilon)$, we introduce the notation 
\begin{equation}
\label{eq:quasi_spherical_tensor}
\mc{T}(\varepsilon) =\sum_{n=0}^\infty \stackrel{n}{\mc{T}}\varepsilon^n\;\;,
\end{equation}
where $\stackrel{0}{\mc{T}}$ is $\mc{T}(\varepsilon)$ evaluated in the background spacetime and $\stackrel{n}{\mc{T}}$ is the $n^{th}$ perturbation of  $\mc{T}(\varepsilon)$ with respect to the background spacetime. 

{With $\mc{R}_{ab}$ denoting the Ricci tensor,  the vacuum Einstein equations $\mc{R}_{ab}=0$ for the line element \eqref{eq:lineelement}, can be grouped into three sets, a group of three so-called  supplementary equations $(\mc{R}_{uu} = 0$ and $\mc{R}_{uA} = 0)$, one trivial equation ($\mc{R}_{ur} = 0$) and six main equations consisting of four hypersurface equations
\begin{subequations}
\begin{eqnarray}
\label{ }
\mc{R}_{rr}(\varepsilon) &=& 0\;\;,\quad
\mc{R}_{rA}(\varepsilon) = 0\;\;,\quad\\
\mc{R}_{(2D)}(\varepsilon) &:=& g^{AB}(\varepsilon)\mc{R}_{AB}(\varepsilon) = 0\;\;,\quad
\end{eqnarray}
\end{subequations}
and two  equations
\begin{equation}
\label{ }
\mc{R}^{TT}_{AB}(\varepsilon):=\mc{R}_{AB}(\varepsilon) - \f{1}{2}g_{AB}(\varepsilon) \mc{R}_{(2D)}(\varepsilon) = 0\;\;.
\end{equation}
This grouping is given by the Bondi--Sachs lemma \cite{Bondietal1962,Sachs1962,TW} obtained from the twice contracted Bianchi identities: {\it If the main equation hold on one null cone and if the optical expansion of null rays\footnote{The outgoing null covector in the Bondi--Sachs coordinates is $k_a=\delta^{u}\!_a{}$ and its  optical expansion is $\nabla_a k^a = (1/r)\exp(-2\beta)$. }  does not vanishes on the cone (\ie $\beta$ is finite), then the trivial equation is fulfilled algebraically and the supplementary equations hold provided they hold at on radius $r> 0$.} Therefore, the supplementary equations can be seen as constraint equations for the metric functions  at given radius $r>0$. As we intent to solve the Einstein equations on the entire null cone including its vertex, the regularity conditions at the vertices can be used to replace these constraints. Although the quasi-spherical approximation being introduced is completely general, we consider for simplicity hereafter only the zeroth and first order terms of in the $\varepsilon$-expansion of the main equations, whose relevant Ricci tensor components are given in App. \ref{app:Ricci}. 
\section{Solution procedure for the background spacetime and linear perturbations}\label{sec:sol_proc}
\subsection{Background spacetime}
\noindent
Here we show that the background spacetime must be Minkowskian, when a regular vertex and a vacuum spacetime are assumed.

From $\stackrel{0}{\mc{R}}_{rr}=0$, $\stackrel{0}{\mc{R}}_{(2D)}=0$ (App. \ref{app:O_0}), and the regularity conditions  \eqref{eq:reg_back} it follows that 
\begin{equation}
\label{eq:sol_beta_Phi_0}
\beta_{0}(x^a)=0 \;\;,\qquad \Phi_0(x^a)=0\;\;.
\end{equation}
Hence the background metric is  Minkowskian  with respect to outgoing null coordinates and has the line element
\begin{equation}
\label{ }
ds^2 = -du^2 -2dudr + r^2 q_{AB}(x^C)dx^A dx^B\;\;.
\end{equation}
Hereafter, we set  $\beta_0=\Phi_0=0$ in all equations and use the standard spherical coordinates $x^A=(\theta,\,\phi)$ to parameterize the unit sphere metric, \ie $q_{AB}(x^A) = \mr{diag}(1,\,\sin^2\theta)$. 
\subsection{Linear perturbations}
\subsubsection{A master equation for vacuum perturbations}
In this section, we derive a covariant differential equation for a master function that allows us to determine all linear perturbations of the Minkowski background spacetime.

From  \eqref{eq:beta_lin} and the regularity conditions  \eqref{eq:reg_pert} it is seen that 
\begin{equation}
\label{eq:sol_beta_1}
\beta_{(1)}(x^a)=0\;\;,
\end{equation}
which is hereafter imposed in the calculations. Setting ${\stackrel{1}{\mc{R}}\!\!_{rA}=0}$ (from eq. \eqref{eq:UA_lin}) yields
\begin{equation}
\label{eq:EE_1_UA}
0= \left(r^4q_{AE}U^E_{(1),r}\right)_{,r} 
    +r^2 q^{EF}D_E \gamma^{(1)}_{AF,r}\;\;.
\end{equation}
From this equation it is clear that if either $\gamma^{(1)}_{AB}$ or $U^A_{(1)}$ is known on the interval $[0,\,\infty)$, eq. \eqref{eq:EE_1_UA} can be used to solve for the other respective field.  
Inserting \eqref{eq:sol_beta_Phi_0} and  \eqref{eq:sol_beta_1}  into \eqref{eq:Phi_lin},  and setting $\stackrel{1}{\mc{R}}_{(2D)}=0$ gives the  equation 
\begin{eqnarray}
\label{eq:EE_1_Phi}
  0&=&     
    \breve{\Phi}_{,r}
      -D^A D^B\gamma^{(1)}_{AB}
  -\f{1}{r^2}D_A(r^4U^A)_{,r}\;\;,
\end{eqnarray}
which allows us to determine  an  intermediate variable $\breve{\Phi}$ that
is related algebraically with $\Phi_{(1)}$ by
\begin{equation}
\label{eq:Phi_intermediate}
\breve{\Phi} = 2r(1+2\Phi_{(1)})\;\;.
\end{equation}
The two equations \eqref{eq:EE_1_UA} and \eqref{eq:EE_1_Phi} link the three perturbation variables. In particular, $U^A_{(1)}$ and $\Phi_{(1)}$ can be calculated once $\gamma^{(1)}_{AB}$ is known  or $\gamma^{(1)}_{AB}$ and $\Phi_{(1)}$ are determined from $U^A_{(1)}$. In what follows we   determine $U^A_{(1)}$  and $\Phi_{(1)}$ from $\gamma^{(1)}_{AB}$.

Defining the functionals  $\ms{A}(\gamma^{(1)}_{AB})$ and $ \ms{B}(D_A X_B)$ acting on $\gamma^{(1)}_{AB}$ and arbitrary covectors $X_A$, respectively, like 
\begin{subequations}
\begin{eqnarray}
\label{ }
 \ms{A}(\gamma^{(1)}_{AB})&:=&
    r(r\gamma^{(1)}_{AB,u})_{,r}-\f{1}{2}\left(r^2\gamma^{(1)}_{AB,r}\right)_{,r}\;\;,\label{eq:def_A}\\
 \ms{B}(D_AX_B)&:=&D_A X_B -\f{1}{2}q_{AB}(q^{EF}D_EX_F) \;\;,  
\end{eqnarray} 
\end{subequations}
allows us to write the evolution equations $\stackrel{1}{\mc{R}}\!\!^{(TT)}_{AB}=0$ (eq.  \eqref{eq:R_TT}) briefly as
\begin{equation}
\label{eq:abs_ev_gam}
0=\ms{A}(\gamma^{(1)}_{AB}) + r^2 \ms{B}\left(q_{AE}D_B U^E_{(1),r}\right)
      +2 \ms{B}\left(q_{AE}D_B U^E_{(1)}\right)     \;\;.
\end{equation}
Taking the covariant derivative $D_B $ of  \eqref{eq:EE_1_UA} we can derive 
 \begin{eqnarray}
0&=&r^2 \ms{B}\Big(q_{AE}D_B U^E_{(1),rr}\Big)
  + 4 r  \ms{B}\Big(q_{AE}D_B U^E_{(1),r }\Big)\nonumber\\
  &&
  + \ms{B}(D_B D_E\gamma^{(1)}_{AE,r}) \;\;.\label{eq:D_eq_UA_1}
\end{eqnarray}
From the following calculation 
\begin{equation}
\df{}{r}\bigg(r^2\left\{\Big[\df{}{r} \eqref{eq:abs_ev_gam} -  \eqref{eq:D_eq_UA_1}\Big]r -  \eqref{eq:abs_ev_gam}\right\}\bigg)+ r^2\eqref{eq:D_eq_UA_1}\;\;,
\end{equation}
we obtain the equation
\begin{eqnarray}
\label{eq:res_manip_gam}
0&=&
  \Big[r^3\ms{A}_{,r}(\gamma^{(1)}_{AB} )
   -r^2\ms{A}_{}\big(\gamma^{(1)}_{AB}\big)\Big]_{,r}
         + r^2\ms{B}\Big(D_AD^E \,\gamma^{(1)}_{BE,r}\Big)\nonumber\\
&& 
         -\bigg[ r^3\ms{B}\Big(D_AD^E \,\gamma^{(1)}_{BE,r}\Big)\bigg]_{,r}.\;\; 
\end{eqnarray}
Defining the auxiliary variable $\chi_{AB}:=r\gamma^{(1)}_{AB}$, using the definition \eqref{eq:def_A} allows us to write \eqref{eq:res_manip_gam} as
\begin{eqnarray}
\label{ }
  0&=&\!\! 
  \Big[ r^4\Big(\chi_{AB,rru}-\f{1}{2}\chi_{AB,rrr}\Big)\Big]_{,r}\!\!
 -r^2 \ms{B}\Big(D_AD^E \,\chi_{BE,rr}\Big)\nonumber\\
\end{eqnarray}%
This a second order differential equation for $\psi_{AB}:=\chi_{AB,rr}$, \ie
\begin{eqnarray}
\label{eq:dgl_psiab}
  0&=& 
   \f{1}{r^2}\Big[r^4(2\psi_{AB,u}-\psi_{AB,r})\Big]_{,r}
    -2D_AD^E \,\psi_{BE}\nonumber\\
   && +q_{AB}\Big(D^ED^F\psi_{EF}\Big)
\;\;.
\end{eqnarray}%
Based on our previous definition of the intermediate variable $\chi_{AB}$, we find the following equation
\begin{equation}
\label{eq:dgl_gAB}
(r\gamma^{(1)}_{AB})_{,rr}=\psi_{AB}\;\;,
\end{equation}
which allows us to determine  $\gamma^{(1)}_{AB}$ from $\psi_{AB}$. 
The remaining field perturbations $U^A_{(1)}$ and $\Phi_{(1)}$ are then obtained by integrating hierachically \eqref{eq:EE_1_UA} and \eqref{eq:EE_1_Phi}.

Since a solution of equation \eqref{eq:dgl_psiab} is the starting point to determine all non-trivial metric fields, we shall call it the {\it master equation} of linearized vacuum perturbations on a null cone. The two tensor field $\psi_{AB}$, which is determined by this differential equation, shall be referred to as the {\it master function} of the linearized vacuum perturbations. By calculating the linearized Riemann tensor with respect to \eqref{eq:lineelement}, we find that  $\psi_{AB}$ determines the  components $\mc{R}_{rArB}$ via 
\begin{equation}
\label{ }
\mc{R}_{rArB} = -\f{1}{2}r\,\psi_{AB}.
\end{equation}
\subsubsection{Representation of the perturbations and their equations in a spin frame work}\label{sec:eth_decomp}
In equations. \eqref{eq:EE_1_UA}, \eqref{eq:EE_1_Phi} and \eqref{eq:dgl_psiab} the angular derivatives are covariant derivatives on a unit sphere. In principle, these covariant derivatives could now be expressed by the corresponding partial ones  utilizing the representation of the unit sphere metric  in terms of the coordinates $\theta$ and $\phi$.  However, we follow the approach of \cite{eth} that became standard \cite{WinicourLRR} when working in numerical relativity with the  Einstein equations in Bondi--Sachs formulation. On the two-surfaces $du=0=dr\, (r>0)$, we introduce a complex dyad $q^A$ and its complex conjugated $\bar{q}^A$ to represent the unit sphere metric $q_{AB}$ and its corresponding covariant derivative $\nabla_A$. The dyad is defined by $q^A q_A=q^A_{,u}=q^A_{,r}=0$  and $q^A \bar{q}_A=q$ with $q>0$. The latter definition covers both commonly used normalizations, the traditional one $q=1$ \cite{eth,N2BMS1966,StewardBook,PenRind}  and the numerical one $q=2$ \cite{WinicourLRR,ethNR,CCE1996},  which is employed in numerical relativity. The unit sphere metric, $q_{AB}$, represented by the dyad is
\begin{equation}
\label{eq:qab_dyad}
q_{AB} =\f{1}{q} \Big(q_{A}\bar{q}_{B}+\bar{q}_Aq_B\Big).
\end{equation}
According to our choice of angular coordinates, $x^A=(\theta,\,\phi)$, $q^A$ can be expressed as $q^A=(q/2)^{1/2}(1,\,i\sin^{-1}\theta)$ with $i=\sqrt{-1}$. 

Any traceless, symmetric tensor $\eta_{(A_1\hdots A_s)}$ of rank $s$ on the two-surfaces $du=0=dr\, (r>0)$ can be expressed by the dyad $q^A$ and a complex scalar field $\eta$ like
\begin{equation}
\label{eq:eta_sym}
\eta_{(A_1\hdots A_s)} = \f{1}{q^s}\Big(\eta \;\overline q_{A_1}\hdots \overline q_{A_s} +\overline \eta \; q_{A_1}\hdots q_{A_s}  \Big)\;\;,
\end{equation}
where
\begin{subequations}
\begin{eqnarray}
 \eta &=& \eta_{(A_1\hdots A_s)} q_{A_1}\hdots q_{A_s}\;\;, \label{eq:det_eta}\\
\overline \eta &=& \eta_{(A_1\hdots A_s)}\overline q_{A_1}\hdots \overline q_{A_s} \;\;.
\end{eqnarray}
\end{subequations}
When the spacelike vectors $\Re e(q^A)$ and $\Im m(q^A)$ are rotated in their complex plane by a real angle $\vartheta$,  the quantity $\eta$ transforms under this rotation as $\eta^\prime = e^{is\vartheta}\eta$ and it is  said to have the spin weight $s$ \cite{N2BMS1966, eth}. For the covariant derivatives $D_B\eta_{(A_1\hdots A_s)} $ with respect to the unit sphere, we define the eth and eth-bar operator as
\begin{subequations}\label{eq:def_eth}
\begin{eqnarray}
\eth \eta&: =& \sqrt{\f{2}{q}} q^{A_1}\hdots q^{A_s} q^B D_B\eta_{(A_1\hdots A_s) }\;\;\\
\overline \eth  \eta&: =& \sqrt{\f{2}{q}}  q^{A_1}\hdots q^{A_s} \overline q^B D_B\eta_{(A_1\hdots A_s) }\;\;,
\end{eqnarray} 
\end{subequations}
which correspond the following derivatives of $\eta$ in terms of the coordinates $x^A$ 
\begin{subequations}
\begin{eqnarray}
\eth \eta & = & (\sin^s\theta)\bigg[ \df{}{\theta}+\f{i}{\sin\theta}\df{}{\phi}\bigg] \bigg(\f{\eta}{\sin^s\theta}\bigg)\;\;,\\
\overline{\eth}\eta & = &\bigg(\f{1}{\sin^s\theta}\bigg)\bigg[ \df{}{\theta}-\f{i}{\sin\theta}\df{}{\phi}\bigg] \big(\eta\sin^s\theta\big)\;\;.
\end{eqnarray}
\end{subequations}
Based on \eqref{eq:def_eth}, we find the commutator of the eth and  eth-bar operator as
\begin{equation}
\label{ }
[\overline \eth, \eth]\eta = 2s \mc{K}\eta\;\;,
\end{equation}
where $\mc{K}=1$ is the Gaussian curvature of the two-surfaces $du=0=dr, (r>0)$ with respect to the unit sphere metric, which is calculated from $\mc{K}:=(1/q^2)q^A\overline q^B q^E \overline q^F\mc{R}_{ABEF}(q_{CD})$.

To write \eqref{eq:EE_1_UA}, \eqref{eq:EE_1_Phi} and \eqref{eq:dgl_psiab} in a spin representation, we define the spin weighted quantities
\begin{subequations}
\begin{eqnarray}
\psi & := & q^Aq^B \psi_{AB} \;\;,\\
\mc{J} & := &q^Aq^B \gamma^{(1)}_{AB}\;\;, \\
\mc{U} & := &q_AU_{(1)}^A\;\;,
\end{eqnarray}
\end{subequations}
where $\mc{U}$ has spin-weight 1 and $\psi$  and $\mc{J}$ have spin-weight 2 \footnote{The linear term in the quasi-spherical expansion of the quantity $J$ in ref. \cite{CCE1996} is related to $\mc{J}$ as $\mc{J}=2J$. }. According to \eqref{eq:det_eta}, the metric perturbations $U^A_{(1)}$ and $\gamma^{(1)}_{AB}$ can be found from $\mc{U}$ and $\mc{J}$ as
\begin{subequations}\label{eq:Ua_gamAB_spin}
\begin{eqnarray}
U^A_{(1)} & = & \f{1}{q}\Big(\mc{U}\overline q^A + \overline{ \mc{U}} q^A\Big) \;\;,\\
\gamma_{AB}^{(1)} & = & \f{1}{q^2}\Big(\mc{J}\,\overline q_A\overline q_B + \overline{ \mc{J}} q_Aq_B\Big) \;\;.\label{eq:gamAB_dyad}
\end{eqnarray}
\end{subequations}
Multiplying 
  \eqref{eq:dgl_psiab} with $q^Aq^B$ yields the spin representation of the master equation
\begin{subequations}\label{spin_pert_eqn_set}
\begin{equation}
\label{eq:dgl_psi}
  0= 
   \f{1}{r^2}\Big[r^4\Big(2\psi_{,u}-\psi_{,r}\Big)\Big]_{,r}
    -\big(\overline{\eth}\eth-4\big)\,\psi
\;\;.
\end{equation}
From the definition of $\psi_{AB}$ we deduce an equation that determines  $\mc{J}$ from $\psi$
\begin{equation}
\label{eq:dgl_psi_gam}
\Big(r\mc{J}\Big)_{,rr} = \psi\;\;.
\end{equation}
Multiplying \eqref{eq:EE_1_UA} with $q^A$ and  using \eqref{eq:qab_dyad} allows us to write \eqref{eq:EE_1_UA} as 
\begin{equation}
\label{eq:EE_1_UA_eth}
0=\Big(r^4\mc{U}_{,r}\Big)_{,r} +\f{ r^2}{\sqrt{2q}}\overline{{\eth}}(\mc{J}_{,r})\;\;. 
\end{equation}
The spin representation of eq. \eqref{eq:EE_1_Phi} can be found as 
\begin{eqnarray}
\label{eq:EE_1_Phi_eth}
  0&=&     
     \breve{\Phi}_{,r}
     - \f{1}{2q}\Big( \eth^2 \bar{ \mc{J}} + \overline{\eth}^2\mc{J}\Big)    \nonumber\\
&&  -\f{1}{r^2\sqrt{2q}}\Big[r^4\Big(\eth \overline{\mc{U}}+\overline{\eth}\mc{U}\Big)\Big]_{,r}\;\;,
\end{eqnarray}
\end{subequations}
where $\breve \Phi$ has a spin weight zero, because it is a tensor of rank zero. Although it is not required hereafter, we give for completeness the evolution equation \eqref{eq:abs_ev_gam} in terms of the spin weighted variables
\begin{eqnarray}\label{eq:EE_1_J}
0&=&
   2(r\mc{J})_{,ur}
      -\f{1}{r}\bigg[ r^2\mc{J} _{,r}
     - r^2(2q)^{1/2}  \eth\, \mc{U}\bigg]_{,r}  \label{eqJ} 
\end{eqnarray}
In equations \eqref{eq:EE_1_UA_eth}, \eqref{eq:EE_1_Phi_eth}, and \eqref{eq:EE_1_J}   it is seen that the $\eth$ and $\overline{\eth}$ operators raise and lower, respectively,  the spin weight when applied  to a spin weighted quantity, since the overall spin weight in these equations must coincide with the spin weight of the variable that does not carry an $\eth$ or $\overline{\eth}$ operator.

Equations \eqref{spin_pert_eqn_set} are the equations  that need to be solved for the perturbations in  the spin representation.
Equation \eqref{eq:dgl_psi} is a wave equation for the spin-2 field $\psi$ and we show in App. \ref{secAPPB} how it is related with the flat -space wave equation of a spin-0 field. The field $\psi$ is, in fact, the linearized Weyl scalar $2r\Psi_0$ \cite{NP1962}. Since  a solution of  \eqref{eq:dgl_psi} determines  all linear perturbations with respect to a Minkowskian background, the perturbations describe the propagation of spin-2 waves on a Minkowskian background spacetime.   
\subsubsection{Decomposition of the perturbations into spin-weighted harmonics}
A standard approach to solve the linearized Einstein equations is to decouple the angular dependence from  the equations by expressing the perturbations in terms of a spectral basis depending on the angular coordinates, only, and where the coefficients of such spectral series depend on all other coordinates but the angles. By inserting such spectral series' into the equations, one then is able to derive differential equations for the coefficients of these series.  Hereafter we follow this approach. 

Since in eqs.  \eqref{eq:dgl_psi}, \eqref{eq:EE_1_UA_eth} and \eqref{eq:EE_1_Phi_eth}  the angular derivatives are expressed in terms of the $\eth$-operator, we  decompose $\psi,\,J,\,U$ and $\breve{\Phi}$ with respect to a basis of spin weighted harmonics $_sY_{lm}$ which are eigenfunctions of the operator $\overline{\eth}\eth$. Let $Y_{lm}(x^A)$ be the  conventional spherical harmonics \cite{Jackson}, then the spin-s weighted spherical harmonics, $_sY_{lm}(x^A)$,  are derived from the  spherical harmonics, $Y_{lm}(x^A)$, like \cite{StewardBook}
\begin{subequations}
\begin{eqnarray}
_sY_{lm} & = & k(l,\,s)\eth^sY_{lm}\;\;\quad\mbox{for}\;\; s>0 \;\;,\\
_sY_{lm} & = & Y_{lm}\;\;\quad\qquad\mbox{for}\;\; s=0\;\;,\\
_sY_{lm} & = & (-)^{|s|}k(l,\,|s|)\overline{\eth}^{|s|} Y_{lm}\;\;\quad\mbox{for}\;\; s<0 \;\;,
\end{eqnarray}
\end{subequations}
where  $k(l,s):=[(l-s)!/(l+s)!]^{1/2}$ and $_sY_{lm}=0$ for $|s|>l$. 
The following properties of the  $\eth$-operator and the $_sY_{lm}$ are used \footnote{Note there is a phase difference in ref. \cite{eth} in the conjugate property (their eq. 2.6) of the $_sY_{lm}$ due to another  phase convention in the definition of the spherical harmonics.}:
\begin{subequations}\label{eq:prop_sYlm}
 \begin{eqnarray}
     _s\overline{Y}_{lm} & = &    (-)^{m+s} \, _{-s}{Y}_{l(-m)} \;\;,\label{eq:eth_conj}\\
     \eth (\,_{s}Y_{lm} )& = &+ \sqrt{(l-s)(l+s+1)}\, _{s+1}Y_{lm}\;\;,\label{eq:eth_up}\\ 
         \overline{ \eth} (\,_{s}Y_{lm}) & = & -\sqrt{(l+s)(l-s+1)}\, _{s-1}Y_{lm}\;\;,\label{eq:eth_dn}\\
         \overline{ \eth} \eth (\,_{s}Y_{lm}) & = & -{(l-s)(l+s+1)}\, _{s}Y_{lm}\;\;,\label{eq:eth_eigenfunction}
\end{eqnarray}
where \eqref{eq:eth_eigenfunction} shows  that $_sY_{lm}$ are the eigenfunctions of the operator $ \overline{ \eth} \eth$ \footnote{\label{ft:eigenfunc_sYlm} The $_sY_{lm}$ are not the only eigenfunctions of $\overline{ \eth} \eth$, there exist in principle another set of eigenfunctions, $_sQ_{lm}(x^A)$ say, which are have the property of being irregular at the poles $\theta\in\{0,\,\pi\}$. }. In particular,  if $s=0$ the operator $\overline{\eth}\eth$ corresponds to the angular momentum operator  since $         \overline{ \eth} \eth Y_{lm}  =  -l(l+1)Y_{lm}$.
\end{subequations}
We make the following ansatz for $\psi,\,\mc{J}$, and $\mc{U}$
\begin{subequations}\label{decomp_psi_J_U}
\begin{eqnarray}
\psi(x^a) & = & \sum_{l=2}^\infty\sum_{m=-l}^l \psi_{lm}(u,r)\,_2Y_{lm}(x^A)\;\;, \label{eq:psi_decomp}\\
\mc{J}(x^a) & = & \sum_{l=2}^\infty\sum_{m=-l}^l J_{lm}(u,r)\,_2Y_{lm}(x^A)\;\;,  \label{eq:J_decomp}\\
\mc{U}(x^a) & = & \sum_{l=1}^\infty\sum_{m=-l}^l U_{lm}(u,r)\,_1Y_{lm}(x^A)\;\;.  \label{eq:U_decomp}
\end{eqnarray}
\end{subequations}
The perturbation variable $\breve{\Phi}$ is a spin-0 field, thus it would be the most natural to express it in a $_0Y_{lm}$ basis. However, an inspection of \eqref{eq:EE_1_Phi_eth} while using \eqref{eq:prop_sYlm} shows that the terms containing $\overline{\eth}^2\mc{J}+\eth^2\overline{\mc{J}}$ and $\overline{\eth}\mc{U}+\eth\overline{\mc{U}}$ have an angular behavior like
\begin{equation*}
\label{ }
 _0Y_{lm}+(-)^m\,_0Y_{l(-m)}\;\;. 
\end{equation*}
By defining the following  spin-0 harmonic
\begin{equation}
\label{eq:def_Zlm}
Z_{lm}(x^A):=\f{1}{2}\Big[\,_0Y_{lm}(x^A)+(-)^m\,_0Y_{l(-m)}(x^A)\Big]\;\;
\end{equation}
and making for $\breve{\Phi}$  the ansatz
\begin{equation}
\breve{\Phi}(x^a):= \sum_{l=1}^\infty\sum_{m=-l}^l \Phi_{lm}(u,r) Z_{lm}(x^A)\;\;,  \label{eq:Phi_decomp}
\end{equation}
 allows us to decouple the angular derivatives in  \eqref{eq:EE_1_Phi_eth}  with  the angular part of $\breve{\Phi}$. 

Inserting \eqref{decomp_psi_J_U} and \eqref{eq:Phi_decomp} into \eqref{spin_pert_eqn_set}, and using \eqref{eq:prop_sYlm} yields an hierachical set of differential equations  coupling the coefficients $\psi_{lm}$, $J_{lm}$, $U_{lm}$, and $\Phi_{lm}$
\begin{subequations}
\begin{eqnarray}
0& = & \f{1}{r^2}\Big[r^4\big(2\psi_{lm,u}-\psi_{lm,r}\big)\Big]_{,r}\nonumber\\
&&+(l+2)(l-1)\psi_{lm}\label{eq:ode_psi_lm}\;, \\
 \Big(r J_{lm}\Big)_{,rr}  & = & \psi_{lm} \label{eq:ode_Jlm}\;,\\
\Big(r^4U_{lm,r}\Big)_{,r} &=& r^2\sqrt{\f{(l+2)(l-1)}{2q}} J_{lm,r}\label{eq:ode_Ulm}\;,\\
\Phi_{lm,r}&=& \f{\sqrt{(l-1)l(l+1)(l+2)}}{q}\,J_{lm}\nonumber\\
&&-\f{1}{r^2}\sqrt{\f{l(l+1)}{2q}}\Big(r^4U_{lm}\Big)_{,r}\label{eq:ode_Philm}\;.
\end{eqnarray}
\end{subequations}
\section{Boundary conditions at the vertex}\label{Sec:RegPsi}
The metric perturbations variables, $\mc{J},\,\mc{U}$, and $\breve{\Phi}$, are functions on null cones $u=const$ with their vertices traced by the central geodesic of a spherically symmetric background spacetime. The null cones, however, are not differentiable at their vertices, consequently the metric perturbation variables are also not differentiable there. As the perturbation $\mc{J}$ defines the complex master function $\psi$, the variable $\psi$ is {\it a priori} also not differentiable at the vertex. In this section we follow the procedure as in \cite{MM2013} to discuss the boundary conditions for $\psi$ and the metric variables at the vertex. Thereby, we assume a Minkowskian observer and the existence of an convex normal neighborhood along the central geodesic of the spherically symmetric background spacetime.

Since the master function $\psi$ is determined from perturbation $\gamma_{AB}^{(1)}$ like 
$\psi =(1/2)(r\gamma_{AB}^{(1)}q^A q^B)_{,rr}$ and since $\gamma_{AB}^{(1)}=O(r^2)$ at $r=0$, we conclude the limiting radial behavior of $\psi$ to be
\begin{equation}
\label{ }
\psi(x^a) = O(r)\;\;.
\end{equation}
To find the exact behavior of $\psi$ near $r=0$, we assume that $\psi$ obeys, at $r=0$, an infinite power series expansion for $\psi$ in terms of $r$ like \begin{equation}
\label{eq:ans_psi_vertex}
\psi(x^a) = \sum^\infty_{n=1} \widetilde{\psi}_n(u, x^A) r^{n}\;\;,
\end{equation}
where the coefficient functions $\widetilde{\psi}_n(u,x^A)$ are assumed smooth functions  for all values of $u$ and $x^A$. 
According to the vertex lemma of \cite{MM2013}, the radial coefficients of \eqref{eq:ans_psi_vertex} must obey the field equations.  Inserting \eqref{eq:ans_psi_vertex} into \eqref{eq:dgl_psi} yields
\begin{eqnarray}       
0 & = & 
    -r\overline{\eth}\eth \psi_1 
    -\sum^{\infty}_{n=1}\bigg\{
    (n+1)(n+4)\widetilde{\psi}_{n+1} \nonumber\\
    &&\quad
       +\big(\overline{\eth}\eth-4\big) \widetilde{\psi}_{n+1}
       -2( n+4)\widetilde{\psi}_{n,u}
          \bigg\}r^{n+1}\;\;.
\end{eqnarray}       
For this equation to be fulfilled, we deduce the conditions 
\begin{subequations} \label{eq:cond_psi}
\begin{eqnarray}
0 & = & \overline{\eth}\eth \psi_1 \label{eq:cond_psi1} \;\;,\\
0 & = &  (n+1)(n+4)\widetilde{\psi}_{n+1} 
       +\big(\overline{\eth}\eth-4\big) \widetilde{\psi}_{n+1}\nonumber\\
       &&
       -2( n+4)\widetilde{\psi}_{n,u} \;\;,\qquad (n\ge 1)\;\;.\label{eq:cond_psin}
\end{eqnarray}
\end{subequations}
To decouple the angular and time dependence, we make the ansatz \footnote{The requirement of a convex normal neighborhood rules out the eigenfunctions $_sQ_{lm}$ of footnote [40], because they are singular at the poles. }
\begin{equation}
\label{eq:decoup_tild_psi_n}
\widetilde{\psi}_n(u, x^A) = \sum^{n+1}_{l=2}\sum_{m=-l}^l c_{l.m.n}(u)\,_2Y_{lm}(x^A)\;\;.
\end{equation}
Inserting \eqref{eq:decoup_tild_psi_n} into \eqref{eq:cond_psi1} yields
\begin{equation}
\label{ }
0 = \overline{\eth}\eth \psi_1 = \sum_{m=-2}^2(-5)(2-2)c_{2.m.1}(u)\,_2Y_{lm}(x^A) \;\;,
\end{equation}
showing that $c_{2.m.1}(u) $ are five functions for the $m-$modes of the $l=2$ spin-2 harmonic, $_2Y_{2m}$, that can be chosen arbitrarily, and which we write as
\begin{equation}
\label{eq:c_21m}
 c_{2.m.1}(u):=C^{(m)}_{2.1}(u)\;\;.
\end{equation}
Inserting \eqref{eq:decoup_tild_psi_n} into \eqref{eq:cond_psin} gives, after little algebra,  for $n\ge1$
\begin{eqnarray}
0 & = &\sum_{m=-(n+2)}^{n+2}c_{n+2.m.n+1}\Big[ 4-4\Big] \,_2Y_{(n+2)m}\nonumber\\
       &&+\sum^{n+1}_{l=2}\sum_{m=-l}^l\bigg\{-2(n+4) \Big[\f{d}{du}c_{l.m.n}\Big]\nonumber\\
       &&\!\!\!\!+c_{l.m.n+1}\Big[ (n+1)(n+4)-(l-2)(l+3)-4\Big]\bigg\} \,_2Y_{lm},\nonumber\\
\end{eqnarray}
from which we conclude that the functions $c_{n+2.m.n+1}(u)$, $(n\ge1)$ are  $2n+5$ arbitrary functions for the $m-$modes of $_2Y_{(n+2)m}$ and the coefficient functions $c_{l.m.n+1}(u)$ are for $(n\ge1)$, $(2\le l \le n+1)$, $|m|\le l$ given by the recursive relation 
\begin{eqnarray}\label{eq:rec_c}
c_{l.m.n+1}(u)&  =&a_{l.n}\f{d}{du}c_{l.m.n}(u)\;\;,\label{eq:c_rek} 
\end{eqnarray}
where 
\begin{equation}
\label{eq:a_ln}
a_{l.n} :=-\f{2(n+4)}{(l+n+3)(l-n-2)}\;\;.
\end{equation} 
Defining the arbitrary functions $c_{n+2.m.n+1}(u)$  as 
\begin{equation}
\label{eq:c_n+2n+1m}
c_{n+2.m.n+1}(u):=C_{n+2.n+1}^{(m)}(u)\;\;,\qquad(n\ge 1)\;\;,
\end{equation}
 and writing out the recursive series \eqref{eq:rec_c} for the first values of $l$ and $n$ allows us to deduce for $(n>1)$, $(2\le l\leq n)$, $|m|\le l$ an explicit form of the recursive series  \eqref{eq:rec_c}
\begin{eqnarray}
\label{eq:c_coff_expl}
c_{l.m.n}(u) &=&\bigg( \prod_{k=l-1}^{n-1}a_{l.k}\bigg)\f{d^{n-l+1}}{du^{n-l+1}}C^{(m)}_{l.l-1}(u)\;\;.\label{eq:c_coeff_expl_text}
\end{eqnarray}
The product term involving the $a_{l.k}$ in \eqref{eq:c_coff_expl} can be simplified further using \eqref{eq:a_ln}  and  properties of the factorial and  Gamma function    
\begin{eqnarray}\label{eq:prod_a}
 \prod_{k=l-1}^{n-1}a_{l.k} &=&\f{2^{n-l}}{(l+1)}\f{(2l+2)!(n+3)!}{(n+l+2)!(n-l+1)!}\;\;.
\end{eqnarray}
Using \eqref{eq:ans_psi_vertex}, \eqref{eq:decoup_tild_psi_n}, \eqref{eq:c_21m},  \eqref{eq:rec_c}, \eqref{eq:c_n+2n+1m}, \eqref{eq:c_coff_expl} and \eqref{eq:prod_a} allows us to find the explicit dependence of  $\psi$ from the arbitrary functions $C_{n.n+1}^{(m)}$, that is
\begin{widetext}
\begin{eqnarray}
\psi (x^a)&=& \sum_{n=1}^\infty \sum_{m=-(n+1)}^{n+1} \Big[C^{(m)}_{n+1.n}(u) \,_2Y_{(n+1)m}(x^A) \Big]r^n\nonumber\\
    &&  +\sum_{n=2}^\infty \sum_{l=2}^{n}\sum_{m=-1}^l \bigg[ 2^{n-l}\,\f{(2l+2)!(n+3)!}{(l+1)(l+2)!(n+l+2)!(n-l+1)!}\bigg]\bigg[\f{d^{n-l+1}}{du^{n-l+1}}C^{(m)}_{l.l-1}(u)\;Y_{lm}(x^A) \bigg]r^n 
    \;.\label{eq:reg_psi}
\end{eqnarray}
\end{widetext}
Equation \eqref{eq:reg_psi} shows  the boundary conditions of $\psi$ in terms of a power series in $r$ at the vertex and functional dependence of $\psi$ from data - the functions $C_{n.n+1}^{(m)}(u)$ - that are given as free functions along the central geodesic. These free functions are the spin-2 multipoles of the complex master function, \ie the Weyl scalar $2r\Psi_0$,  which are calculated with respect to a Fermi observer following the central geodesic. 
Given $\psi$ we find the boundary conditions for the perturbation $\mc{J}, \,\mc{U},$ and $\Phi_{(1)}$ as  
\begin{widetext}
\begin{subequations}\label{eq:reg_BS}
      \begin{eqnarray}
 \mc{J} (x^a)&=& \sum_{l=2}^\infty \sum_{m=-l}^l \Big[I^{(m)}_{n.n}(u) \,_2Y_{nm}(x^A) \Big]r^n 
  +\sum_{n=3}^\infty \sum_{l=2}^{n-1}\sum_{m=-l}^l b(n,\,l)\bigg[\f{d^{n-l}}{du^{n-l}}I^{(m)}_{l.l}(u)\;_2Y_{lm}(x^A) \bigg]r^{n}     \;\;,
    \label{eq:J_reg_full}
          \end{eqnarray}
                \begin{eqnarray}
\mc{U}(x^a)&=& \sum_{l=2}^\infty \sum_{m=-l}^l \Big[\f{l}{(1-l)(l+2)}\f{\sqrt{(l-2)(l+3)}}{\sqrt{2q}}\,I^{(m)}_{l.l}(u) \,_1Y_{lm}(x^A) \Big]r^{l-1} \nonumber\\
 && +\sum_{n=3}^\infty \sum_{l=2}^{n-1}\sum_{m=-l}^l \f{n\, b(n,\,l)}{(1-n)(n+2)}\f{\sqrt{(l-2)(l+3)}}{\sqrt{2q}}\bigg[\f{d^{n-l}}{du^{n-l}}I^{(m)}_{l.l}(u)\;_1Y_{lm}(x^A) \bigg]r^{n-1}  \;\;,
   \label{eq:U_reg_full}
             \end{eqnarray}
                \begin{eqnarray}
\Phi_{(1)}(x^a)&=& \sum_{l=2}^\infty \sum_{m=-l}^l \Big[\f{\sqrt{(l-1)l(l+1)(l+2)}\,I^{(m)}_{l.l}(u) \,Z_{lm}(x^A)}{4q(l+1)} \Big]r^l\nonumber \\
&&  +\sum_{n=3}^\infty \sum_{l=2}^{n-1}\sum_{m=-l}^l \f{\sqrt{(l-1)l(l+1)(l+2)}\,b(n,\,l)}{4q(n+1)}\bigg[\f{d^{n-l}}{du^{n-l}}I^{(m)}_{l.l}(u)\;Z_{lm}(x^A) \bigg]r^{n}   -\f{1}{2}\;\;,  \label{eq:phi_reg_full}
\end{eqnarray}
\end{subequations}
\end{widetext}
where we defined
\begin{eqnarray}
\label{ }
I_{n.n}^{(m)}(u)&:=&\f{C_{n.n-1}^{(m)}(u)}{n(n+1)}\;\;,\\
b(n,l)&:=&2^{n-1-l}\,\f{l(n+2)(2l+2)!(n-1)!}{(l+2)!(n+l+1)!(n-l)!}\;.
\end{eqnarray}
It is seen in \eqref{eq:J_reg_full} that for static spacetimes the free data $I^{(m)}_{l.l}(u)$ of the $l-$multipole  the along the central geodesic correspond with power $r^l$, and for dynamical spacetimes the $n^{th}$ time derivative of the $l-$multipole of the free data are found at the power $r^{n+l}$. 
Moreover, formula \eqref{eq:J_reg_full} agrees in axial symmetry (\ie $m=$0) with expressions given in \cite{MM2013}. 

\section{Solutions approaches to  the master equation}\label{sec:2sols}
In this section,  we present two different approaches to solve the differential equation for the spectral coefficients of the   complex master function $\psi$. In the first approach (Sec. \ref{sec:1approach}), we make a `standard'  ansatz to solve the differential equation, whereas in the second approach (Sec. \ref{sec:2approach}) we impose an ansatz based on the characteristic nature of the partial differential equation. In both approaches we look for solutions which respect the regularity conditions of the previous sections and we also provide the initial data for the free functions $C_{n+1.n}^{(m)}(u)$. 
\subsection{First Approach }\label{sec:1approach}
To solve the differential equation \eqref{eq:ode_psi_lm} for the coefficients of $\psi_{lm}$, we make a `standard' ansatz of separation of variables 
\begin{equation}
\label{eq:ans_1}
\psi_{lm}(u,r)=e^{T(u)}R_{lm}(r)\;\;.
\end{equation}
Inserting \eqref{eq:ans_1} into \eqref{eq:ode_psi_lm} yields 
\begin{equation}
\label{eq:spe_ans_1}
\f{dT}{du}= \f{r^2\f{d^2R_{lm}}{dr^2}+4r \f{dR_{lm}}{dr} - \Big[ l(l+1)-2\Big]R}{8rR_{lm} +2r^2\f{dR_{lm}}{dr}}\;\;. 
\end{equation}
Since the right hand side of \eqref{eq:spe_ans_1} is independent of $u$, it can be treated as a constant in the $u-$integration that we set to be $dT/du=B=const$. This  implies the solution $T(u) = Bu +u_0$, where the constant $u_0$ is determined by the initial conditions, and  without loss of generality we may set $u_0=0$. Inserting $dT/du=B$ into \eqref{eq:spe_ans_1}   and rearranging  gives 
\begin{equation}
\label{eq:sep_R}
0= r^2\f{d^2R_{lm}}{dr^2}+\Big(4r-2r^2B \Big)\f{dR_{lm}}{dr} - \Big(8rB  + l(l+1)-2\Big)R_{lm}\;\;. 
\end{equation}
To find a solution for \eqref{eq:sep_R},  we make the ansatz $R_{lm}(r)=r^k e^{ar}A_{lm}(r)$  and obtain
\begin{eqnarray}
0 & = & r^2\f{d^2A_{lm}(r)}{dr^2}+\Big[2(a-B)r^2 + 2(k+2)r \Big]\f{dA_{lm}(r)}{dr}  \nonumber\\
 & & \Big[ a(a-2B)r^2 + (4a+2ka-2Bk-8B)r\Big]A_{lm}(r)\nonumber\\
 &&-l(l+1)+2-3k-k^2\Big]A_{lm}(r)\;\;.\label{eq:A_ans}
\end{eqnarray}
If we choose $k=-1$,  $a=B$,  and make in \eqref{eq:A_ans} the variable transformation $z=B r$, \eqref{eq:A_ans} can be cast into a inhomogeneous  Bessel type differential equation 
\begin{equation}
\label{eq:ode_Az}
z^2\f{d^2A_{lm}}{dz^2} + z \f{dA_{lm}}{dz} -\Big[z^2+l(l+1)\Big]A_{lm}=4zA_{lm}\;\;,
\end{equation}
where the inhomogenity is given by the right hand side of \eqref{eq:ode_Az}. The homogeneous counterpart to \eqref{eq:ode_Az} is  the Bessel differential equation 
\begin{equation}
\label{eq:ode_hom_Bessel}
z^2\f{d^2A_{lm}(z)}{dz^2} + z \f{dA_{lm}(z)}{dz} -\Big[z^2+l(l+1)\Big]A_{lm}(z)=0\;\;,
\end{equation}
whose solutions are the modified spherical Bessel functions, $i_l(z)$ and $k_l(z)$, of first and second kind, respectively, 
\begin{equation}
\label{ }
i_{l}(z):=\sqrt{\f{\pi}{2z}}I_{l+\f{1}{2}}(z)\;\,,\;\;
k_{l}(z):=\sqrt{\f{\pi}{2z}}K_{l+\f{1}{2}}(z)\;\;,
\end{equation} 
where $I_{l+\f{1}{2}}(z)$ and $K_{l+\f{1}{2}}(z)$ are the modified Bessel functions of first and second kind \cite[p. 437]{AbramSteg}. The regularity conditions \eqref{eq:reg_psi} for $\psi$ require that $R_{lm}(r)=O(r^{l-1})$ for $r\approx 0$ and $l\ge2$, consequently $A_{lm}(z)$ must behave as $O(z^l)$ for $z\approx 0$. Since the functions $i_l(z)$ and $k_l(z)$ behave as $O(z^l)$ and $O(z^{-l})$ for $z\approx 0$, respectively, the modified  spherical Bessel functions of the second kind must be ruled in the construction of a  regular solution to \eqref{eq:ode_Az}. 
To solve \eqref{eq:ode_Az}, we make the ansatz

\begin{eqnarray}
\label{eq:ans_Al}
A_{lm}(z) &=&  \sum_{k=0}^2  z^k\bigg\{e^{(lm)}_k  i_{l}(z) +   f^{(lm)}_k   i_{l-1}(z)\bigg\}\;\;.
\end{eqnarray}
 where are the coefficients $e^{(lm)}_k$, and $f^{(lm)}_k$, are determined by the boundary conditions at $z=0$ and by inserting \eqref{eq:ans_Al} into \eqref{eq:ode_Az}. Inserting  \eqref{eq:ans_Al} into \eqref{eq:ode_Az} allows us to deduce the solution
 
\begin{eqnarray}
\label{ }
A_{lm}(z) &=&  f^{(lm)}_2\big(z^2+z\big) i_{l}(z) \nonumber\\
&&+ f^{(lm)}_2  \Big[z^2+ (1-l)z+\f{l(l-1)}{2}\Big]i_{l-1}(z)\;\;.\nonumber\\
\end{eqnarray}
Thus a master function  $\psi$ with the following spectral coefficients
\begin{eqnarray}
\lefteqn{\psi_{lm}(u,r)= \f{f^{(lm)}_2}{r}e^{B(u+r)}\bigg\{\Big[(Br)^2+Br\Big]i_{l-1}(Br)}\nonumber\\
 &&+\Big[(Br)^2-(l-1)(Br)+\f{l(l-1)}{2}\Big]i_{l}(Br)\bigg\}\label{eq:psi_bessel}
\end{eqnarray}
respects the regularity condition \eqref{eq:reg_psi} at the vertex. In particular, if we choose in \eqref{eq:reg_psi} the free initial data to be 
\begin{equation}
\label{eq:free_data_bessel}
C^{(m)}_{l+1.l}(u)= f_2^{(lm)}\bigg[\f{1}{(2l+1)!!}+\f{l(l-1)}{2(2l+3)!!}\bigg] e^{Bu} B^l\;\;,
\end{equation} 
where $l!!$ is the double factorial, then the free data \eqref{eq:free_data_bessel}  describe the solution \eqref{eq:psi_bessel} in a neighborhood of $r=0$.

It  can be easily seen in \eqref{eq:psi_bessel} that the spectral coefficients $\psi_{lm}$  diverge exponentially as $r$ tends to infinity. Consequently the spectral coefficients of  linear perturbations that are calculated from  \eqref{eq:psi_bessel} will also diverge exponentially and hence the solution which we found for the master functions will not give to an asymptotically flat spacetime.

\subsection{Second Approach}\label{sec:2approach}
In this section, we present an approach  yielding  a solution of the master equation that is regular at the vertex and finite, when $r$ tends to infinity.

Defining $\psi_{lm} = (1/r^4)\int r^4 P_{lm}dr$ shows that   \eqref{eq:ode_psi_lm} is an integro-differential transport equation
\begin{eqnarray}
 \!0  &= &
   P_{lm,u}-\f{1}{2}P_{lm,r}
   +\f{l(l+1)-6}{2r^6}\,\int r^4 P_{lm} dr
\;.\label{eq:idt_w_lm}
\end{eqnarray}
It can be seen that the integral part of  \eqref{eq:idt_w_lm} vanishes for $l=2$ implying  the  surfaces $u+2r=const$ to be the (ingoing) characteristic surfaces of this integro-differential transport equation.

Since eq. \eqref{eq:ode_psi_lm} is  another representation of the transport equation \eqref{eq:idt_w_lm} and based on the fact that ${u+2r}$ is a characteristic surface for the lowest ( $l=2$) multipole of $\psi$, we make the following ansatz to solve \eqref{eq:ode_psi_lm} 
\begin{equation}
\label{eq:ans_char}
   \psi_{lm} = C_{lm} u^i r^j (u+2r)^k\;\;,
\end{equation}
where the exponents $i,\,j,\,k$ need to be determined by inserting \eqref{eq:ans_char} in \eqref{eq:ode_psi_lm} and $C_{lm}\in\mathbb{R}$ are arbitrary constants. The calculation yields the following possibilities for the exponents for $(l\le 2)$
\begin{subequations}
\begin{eqnarray}
[i,\,j,\,k] &=& \Big[(l+2),\, -(l+2),\,(l-2)\Big]\;\;,\label{eq:coeff_1stposs}\\ 
\,[i,\,j,\,k] &=& \Big[-(l-1),\,(l-1),\,(l+3)\Big]\;\;.\label{eq:coeff_2ndposs}
\end{eqnarray}
\end{subequations}
By inserting these values into \eqref{eq:ans_char} and expanding the thus obtained function at $r=0$, it is seen that the first possibility, \eqref{eq:coeff_1stposs}, for the exponents gives rise to a singular behavior of the coefficients $\psi_{lm}$ in \eqref{eq:ans_char} at ${r=0}$, whereas the second one complies with the regularity requirement \eqref{eq:reg_psi}. Hence,  $\psi$ with following spectral coefficients 
\begin{equation}
\label{eq:psilm_reg}
   \psi_{lm}(u,r) = C_{lm} \f{r^{l-1}}{u^{l-1}(u+2r)^{l+3}}\;\;,
\end{equation}
yields for $u>0$ a  complex master function $\psi$ that is regular at the vertex at $r=0$ and also finite when $r\rightarrow\infty$. The initial data along the central geodesic for $u>0$ are 
\begin{equation}
\label{ }
C^{(m)}_{l+1.l} = \f{C_{lm}}{u^{2(l+2)}}\;\;, \qquad l\geq 1\;\;,\quad |m|\le l\;\;.
\end{equation}
\section{Asymptotically flat and regular solution for the vacuum perturbations}\label{sec:RegSol}
Given the spectral coefficients \eqref{eq:psilm_reg} for the complex master function and the boundary conditions for the perturbations \eqref{eq:reg_pert}, we now integrate \eqref{eq:ode_Jlm}, \eqref{eq:ode_Ulm}, and \eqref{eq:ode_Philm} for the spectral coefficients of the perturbations. Together with $\beta_{(1)}=0$ and \eqref{eq:Phi_intermediate}, our solution of the linearized perturbations $\mc{J},\,\mc{U},$ and $\Phi_{0}$ in term of spin-weighted harmonics are 
\begin{widetext}
\begin{subequations}\label{eq:sol_SPIN2}
\begin{eqnarray}
\mc{J}(x^a)& = &\sum_{l=2}^\infty \sum_{m=-l}^l \f{C_{lm}}{l(l+1)(l+2)} \f{r^l [(l+2)u+4r]}{u^{l+2}(u+2r)^{l+1}}\,_2Y_{lm}(x^A) \;\;,\\
\mc{U}(x^a)&=&\f{1}{\sqrt{2q}}\sum_{l=2}^\infty \sum_{m=-l}^l  \f{C_{lm}}{l(l+1)\sqrt{(l-1)(l+2)}}\,\f{r^{l-1}(lu+2r)}{u^{l+3}(u+2r)^{l}}\,_1Y_{lm}(x^A)\;\;,\\
\Phi_{(1)}(x^a)&=&-\f{1}{2}-\f{1}{4q}\sum_{l=2}^\infty \sum_{m=-l}^l \f{C_{lm}}{\sqrt{(l-1)l(l+1)(l+2)}}\f{r^{l}}{u^{l+3}(u+2r)^{l-1}} \,Z_{lm}(x^A)\;\;,
\end{eqnarray}
\end{subequations}
\end{widetext}
which is well defined for $u>0$, and where the coefficients $C_{lm}$ are arbitrary constants. The  solution above describes for $u>0$  spin-2 fields propagating on a Minkowski background spacetime, and we shall therefore refer to it hereafter as SPIN-2. 

\subsection{Asymptotic properties}
\subsubsection{Conformal compactification, frame at null infinity}
The study of the asymptotic behavior of solutions of the Einstein field equations  \footnote{Bondi, Sachs and collaborators originally studied the  asymptotic behavior of the gravitational fields by  considering expansions of the metric variables in terms of inverse powers of radial coordinate $r$.} is most elegantly done by using Penrose's definition of asymptotically flat spacetime \cite{Penrose1963}. The idea of this compactification is that, aside from a ``physical manifold'' $\ms{M}$ with a ``physical metric'' $g_{ab}$, there exists a positive function $\ell$, which decreases along all complete null geodesics, approaching to zero as their affine parameter goes to infinity. Thereby a ``non-physical metric'' $\hat{g}_{ab}=\ell^2 g_{ab}$ can be extended smoothly to a larger, compactified manifold $\widehat{M} = \ms{M}\cup\partial \ms{M}$ \cite{Friedrich1988}. The boundary $\ms{I}:=\partial\mc{M}$ is called {\it null infinity} and one has $\ell=0$ and $\nabla_a \ell\neq0$ on $\ms{I}$. Points on $\ms{I}$ in the manifold $\widehat{\ms{M}}$ correspond to ``points at infinity" for radiative fields in the physical manifold. It can be shown that  the boundary $\ms{I}$ is a null hypersurface, and that the Weyl tensor, $\hat{\mc{C}}_{abcd}$, behaves as $O(\ell)$ in the neighborhood of $\ms{I}$  \cite{Penrose1963}.  

To find a compactified metric $\hat{g}_{ab}$ for a given Bondi--Sachs metric of a physical spacetime, we have to find a conformal factor $\ell$ such that it has the properties at $\ms{I}$ as stated above.   Assuming $g_{ab}$ is smooth in $\ms{M}$, we first define coordinates $\hat{x}^a$ as a function of the Bondi--Sachs coordinates $x^a$ such that `points at $r=\infty$' in the Bondi--Sachs coordinates are located at finite values in $\hat{x}^a$ coordinates.  Second we calculate an intermediate metric, $g^{*}_{ab}$ say,  by transforming the  Bondi--Sachs metric to the conformal coordinates $\hat{x}^a$ and finally we find a conformal factor $\ell$ making  $\ell^2 g^*_{ab}$ finite at the location of null infinity.  The coordinates $\hat{x}^a$ are called conformal Bondi--Sachs coordinates. 

The coordinate $\hat{x}^{0}:=u$ and the coordinates $\hat{x}^A:=x^A$ are defined as their counterpart in the Bondi--Sachs coordinates.   The coordinate $\hat{x}^1:=\hat{x}$  ranges over the interval $[0,\,a],\, a>0$, and is connected to the  physical radial coordinate $r$ via a  strictly monotonous positive function $r=r(\hat{x})$  having  the further properties $r(0)=0$ and $\lim_{\hat{x}\rightarrow a}r(\hat{x})=\infty$. The points at null infinity $\ms{I}$ are located at $\hat{x}^a=(\hat{u},\,a,\,\hat{x}^A)$ where $\hat{u}$ and $\hat{x}^A$ are arbitrary.  The requirement $\lim_{\hat{x}\rightarrow a}r(\hat{x})=\infty$ and the monotonicity imply that $r(\hat{x})$ has a singularity at $\hat{x}=a$ which is a pole. 
Hence the most general form of $r(\hat{x})$ may be assumed to be of the form 
\begin{equation}
\label{eq:gen_rx}
r(\hat{x}) = \f{R(\hat{x})}{(a-\hat{x})^m}\;\;,
\end{equation}
where $m\in\{1,\,2,\,...\}$ is the order of the pole of $r(\hat{x})$ and $R(\hat{x})$ is a strictly monotonous positive function with $R(0)=0$ and $R(a)\neq0$. If one now transforms the metric components of the  line element \eqref{eq:lineelement} to conformal coordinates $\hat{x}^a$ and assumes that metric functions $h_{AB},\,\beta,\,U^A, $ and $\Phi$ have only poles at $\hat{x}=a$, it can be seen that the thus obtained metric $g^*_{ab}$ has a poles of the order $2m$ and $m+1$. Assuming a conformal factor like $\ell=(a-x)^{k/2},\,k=\{1,\,2,...\}$ and requiring $\hat{g}_{ab}=\ell^2g^*_{ab}$ to be finite as $x\rightarrow a$ implies 
\begin{equation}
\label{ }
k=2\;\;,\qquad m=1.
\end{equation}
Thus the function $\ell(\hat{x}) = a-\hat{x}$ vanishes for $\hat{x}=a$ and its derivative does not vanish at $\hat{x}=a$. Therefore the surface with the coordinate values  $\hat{x}^a=(\hat{u},\,a,\hat{x}^A)$, with arbitrary values of $\hat{u},\,$ and $\hat{x}^A$, corresponds  to null infinity.

The conformal metric $\hat{g}_{ab}$ has the non-trivial covariant components
\begin{subequations}\label{eq:BS_conf}
\begin{eqnarray}
\hat{g}_{\hat{u}\hat{u}} & = & -(a-\hat{x})^2 e^{2\Phi+4\beta} + R^2(\hat{x})h_{\hat{A}\hat{B}} U^{\hat{A}} U^{\hat{B}} \\
\hat{g}_{\hat{u}\hat{x}} & = & -\Big[(a-\hat{x})\f{dR}{d\hat{x}}+R(\hat{x}) \Big] e^{2\beta} \\
\hat{g}_{\hat{u}\hat{A}} & = & -R^2(\hat{x})\hat{h}_{AB}\hat{U}^{{B}} \\
\hat{g}_{\hat{A}\hat{B}} & = & R^2(\hat{x})\hat{h}_{AB} \;\;.
\end{eqnarray}
\end{subequations}
and the nonzero contravariant components
\begin{subequations}\label{eq:BS_conf}
\begin{eqnarray}
\hat{g}^{\hat{x}\hat{u}} & = & -\Big[(a-\hat{x})\f{dR}{d\hat{x}}+R(\hat{x}) \Big]^{-1} e^{-2\beta} \;\;,
\end{eqnarray}
\begin{eqnarray}
\hat{g}^{\hat{x}\hat{x}} & = & -\f{(a-\hat{x})^2 e^{2\Phi}}{(a-\hat{x})\f{dR}{d\hat{x}}+R(\hat{x})}\;\;,\\
\hat{g}^{\hat{x}\hat{A}} & = & \hat{U}^A\Big[(a-\hat{x})\f{dR}{d\hat{x}}+R(\hat{x}) \Big]^{-1} e^{-2\beta} \\
\hat{g}^{\hat{A}\hat{B}} & = & \f{1}{R^2(\hat{x})}\hat{h}^{AB} \;\;.
\end{eqnarray}
\end{subequations}
In the quasi-spherical approximation of SPIN-2, the conformal metric $\hat{g}_{ab}$ is expanded in terms of the $\varepsilon-$parameter as
\begin{equation}
\label{ }
{\hat{g}}_{ab} \hat{=} \stackrel{0}{\hat{g}}_{ab}+\varepsilon \stackrel{1}{\hat{g}}_{ab}\;\;,
\end{equation}
where $ \stackrel{0}{\hat{g}}_{ab}$ has the non-vanishing components
\begin{subequations}\label{eq:background_scri}
\begin{eqnarray}
\stackrel{0}{\hat{g}}_{\hat{u}\hat{u}} & = &-(a-\hat{x})^2\;\;, \\
\stackrel{0}{\hat{g}}_{\hat{u}\hat{x}}& = &-\Big[(a-\hat{x})\f{dR}{d\hat{x}}+R(\hat{x}) \Big]\;\;, \\
\stackrel{0}{\hat{g}}_{AB} & = & R^2(\hat{x}){q}_{AB} \;\;.\label{eq:gAB_conf}
\end{eqnarray}
\end{subequations}
The non-zero components of $\stackrel{1}{\hat{g}}_{ab}$ are 
\begin{subequations}\label{eq:pert_scri}
\begin{eqnarray}
\stackrel{1}{\hat{g}}_{\hat{u}\hat{u}} & = & -2(a-\hat{x})^2\hat{\Phi}_{(1)} \;\;, \\
\stackrel{1}{\hat{g}}_{\hat{u}\hat{x}}  & = & -\Big[(a-\hat{x})\f{dR}{d\hat{x}}+R(\hat{x}) \Big] \;\;, \\
\stackrel{1}{\hat{g}}_{\hat{u}\hat{A}}  & = & -R^2(\hat{x}){q}_{AB}\hat{U}^{B}_{(1)}\;\;, \\
\stackrel{1}{\hat{g}}_{\hat{A}\hat{B}}  & = & R^2(\hat{x})\hat{\gamma}^{(1)}_{AB} \;\;,
\end{eqnarray}
\end{subequations}
where $\hat U^A_{(1)}$ and $\hat \gamma^{(1)}_{AB}$ are calculated from $\hat{\mc{U}}$ and $\hat{\mc{J}}$ using \eqref{eq:Ua_gamAB_spin}.

If we calculate $\stackrel{0}{\hat{g}}_{ab}$ at null infinity, \ie set $\hat{x}=a$, and define new coordinates $\tilde{x}^a$ as
\begin{equation}
\label{ }
\tilde{x}^0:=\tilde{u}=\hat{u}\;\;,\quad
\tilde{x}^1:=\tilde{x}=-R(a)\hat{ x}\;\;,\quad
\tilde{x}^A:=\tilde{x}^A=\hat{x}^A\;\;,
\end{equation}
then a coordinate transformation of \eqref{eq:background_scri}  at $\ms{I}$ from $\hat{x}^a$ to $\tilde{x}^a$ brings  \eqref{eq:background_scri} to a conformal Minkowski metric $\tilde{g}_{ab}$ with the non-zero components
\begin{eqnarray}
\label{eq:BondiFrame}
\tilde{g}_{\tilde{u}\tilde{x}}\Big|_{\ms{I}}(\tilde{x}^a) = 1\;\;,\qquad
\tilde{g}_{AB}\Big|_{\ms{I}}(\tilde{x}^a) = q_{AB}\;\;,\qquad
\end{eqnarray}
which corresponds to  the metric of an inertial observer  in the (compactified) conformal spacetime  $(\widehat{\mc{M}}, \hat g_{ab})$\cite{TW}. The inertial conformal  frame is the frame at null infinity where one can uniquely define the asymptotic properties of an asymptotically flat spacetime, because most generally a frame at null infinity is not a Minkowskian one \cite{GerochHorowitz}. The Bondi--Sachs variables have in an inertial frame at null infinity the values
\begin{equation}
\label{ }
\tilde{h}_{AB}|_{\ms{I}} = q_{AB} \;\;,\quad 0=\tilde{U}^A|_{\ms{I}}=\tilde{\beta}|_{\ms{I}}=\breve{\Phi}|_{\ms{I}}\;\;,
\end{equation}
which means that the perturbations $\beta_{(1)},\,\mc{U},\,\mc{J},$ and $\Phi_{(1)}$   have the values
\begin{equation}
\label{ }
 0=\tilde{\mc{J}}|_\ms{I}=\tilde{\mc{U}}|_{\ms{I}}=\tilde{\beta}_{(1)}|_{\ms{I}}=\tilde{\Phi}_{(1)}|_{\ms{I}}\;\;,
\end{equation}
Inserting \eqref{eq:gen_rx} into \eqref{eq:sol_SPIN2} and expanding the thus obtained expression at $\hat{x}=a$ yields  
\begin{subequations}\label{eq:exp_Spin2_scri}
\begin{widetext}
  \begin{eqnarray}
    \hat{\mc{J}}(\hat{x}^a) & = & \sum_{l=2}^\infty \sum_{m=-l}^l \f{C_{lm}\,_2Y_{lm}(\hat{x}^A)}{2^{l-1} \hat{u}^{l+2}l(l+1)(l+2)}
           +O\Big[(a-\hat{x})\Big] \;\;,\label{eq:exp_J_I}\\
     \hat{\mc{U}}(\hat{x}^a)&=&    
          \f{1}{\sqrt{2q}} \sum_{l=2}^\infty \sum_{m=-l}^l \f{C_{lm}\,_1Y_{lm}(\hat{x}^A)}{2^{l-1} \hat{u}^{l+3}\,l(l+1)\sqrt{(l-1)(l+2)}}
           +O\Big[(a-\hat{x})^2\Big]\;\;,\label{eq:exp_U_I}\\
     \hat{\Phi}_{(1)}(\hat{x}^a)&=& -\f{1}{q} \sum_{l=2}^\infty \sum_{m=-l}^l \f{R(a)C_{lm} Z_{lm}(\hat{x}^A)}{(a-\hat{x})2^{l+1}u^{l+3}\sqrt{(l-1)l(l+1)(l+2)}}+ O\Big[(a-\hat{x})^0\Big]\label{eq:exp_Phi_I}
     \;\;.
\end{eqnarray}    
\end{widetext}
\end{subequations}
Expressions \eqref{eq:exp_Spin2_scri} show that  $\hat{\mc{J}},\,\hat{\mc{U}}$, and $\hat{\Phi}_{(1)}$   evaluated at $\ms{I}$ exhibit non-trivial values which indicate that SPIN-2 does not result in an inertial  conformal  frame at null infinity after applying  Penrose compactification.  Note although \eqref{eq:exp_Phi_I} diverges at $\ms{I}$, the corresponding metric components $g_{\hat{u}\hat{u}}$ and $g^{\hat{x}\hat{x}}$ are finite at $\ms{I}$.
\subsubsection{On the determination of $\Psi_4$ in a conformal frame}
To extract unambiguously  physical information from SPIN-2 at null infinity, we have to find, in principle,  a coordinate transformation that casts the metrics \eqref{eq:background_scri} and \eqref{eq:pert_scri} into a conformal Bondi frame at null infinity.  In this section, we determine the Weyl scalar $\Psi_4$ \cite{Penrose1963,NP1962} in a Bondi frame at $\ms{I}$ by following \cite{Babiuc2009}. In this procedure, we do not look for the exact coordinate transformation from the conformal frame to a Bondi frame, but start with an expression of $\Psi_4$ in a Bondi frame and then transform it into an arbitrary conformal frame at $\ms{I}$. We begin with a summary and motivation of the basic ideas of \cite{Babiuc2009}   
 and give the expression of  $\Psi_4$ in a Bondi frame at $\ms{I}$ for the solution SPIN-2 in the end. 

From the conformal metric $\widehat{g}_{ab} = \ell^2g_{ab}$ in $\widehat{\ms{M}}$ and the vanishing Ricci tensor, $\mc{R}_{ab}(g_{ab})$,  in the physical manifold $\ms{M} = \widehat{\ms{M}}/\ms{I}$ the following equation can be derived
\begin{equation}
\label{eq:conf_ric}
0=\ell^2 \widehat{\mc{R}}_{ab} +2\ell \widehat{\nabla}_a\widehat{\nabla}_b\ell + \widehat{g}_{ab}\bigg[\widehat{\nabla}^c\widehat{\nabla}_c\ell -3(\widehat{\nabla}^c\ell)(\widehat{\nabla}_c\ell)\bigg]. 
\end{equation}
Taking the trace of \eqref{eq:conf_ric} and subsequently the limit $(\ell\rightarrow 0)$ of this trace, shows that the surface $\ms{I}$ with the normal vector $\widehat{\nabla}_c\ell$ is a null hypersurface and that $(\widehat{\nabla}^c\ell)(\widehat{\nabla}_c\ell) = O(\ell)$ . From \eqref{eq:conf_ric} we can derive two equations
\begin{subequations}\label{eq:mod_conf_ric}
\begin{eqnarray}
0 & = & \ell \widehat{\mc{R}} + 6\widehat{\nabla}^a\widehat{\nabla}_a \ell - \f{12}{\ell}(\widehat{\nabla^a\ell})(\widehat{\nabla}_a\ell) \;\;, \label{eq:mod_conf_ric_1}\\
0 & = & \ell\bigg(\widehat{\mc{R}}_{ab}-\f{1}{4}\widehat{g}_{ab}\widehat{\mc{R}}\bigg)+2\bigg[\widehat{\nabla}_a\widehat{\nabla}_a\ell - \f{1}{4}\widehat{g}_{ab}(\widehat{\nabla}^c\widehat{\nabla}_c\ell)\bigg]\;,\nonumber\\ \label{eq:mod_conf_ric_2}
\end{eqnarray}
\end{subequations}
where $\widehat{\mc{R}} $ is the Ricciscalar with respect $\widehat{g}_{ab}$ and we used \eqref{eq:mod_conf_ric_1} to eliminate the term $(\widehat{\nabla}^c\ell)(\widehat{\nabla}_c\ell)$ in \eqref{eq:conf_ric} to obtain \eqref{eq:mod_conf_ric_2}. 
Evaluating \eqref{eq:mod_conf_ric} at $\ms{I}$, allows us to define two fields that are intrinsic to  $\ms{I}$ 
\begin{subequations}

\begin{eqnarray}
\widehat{\Theta} & := &\widehat{\nabla}^a\widehat{\nabla}_a \ell \;\; ,\\
\widehat{\Xi}_{ab} &: = & \ell \widehat{\Sigma}_{ab}  \;\;,
\end{eqnarray}
\end{subequations}
where 
\begin{equation}
\label{eq:def_sigma}
\widehat{\Sigma}_{ab}:=\widehat{\nabla}_a\widehat{\nabla}_a\ell - \f{1}{4}\widehat{g}_{ab}(\widehat{\nabla}^c\widehat{\nabla}_c\ell)
\end{equation}
and which have been first found in \cite{TW}, and are crucial in the calculation of $\Psi_4$ in \cite{Babiuc2009}. Using $\hat{\ell} = (a-\hat{x})$, we find  $\hat{\Theta}$ as
\begin{eqnarray}
\hat{\Theta} & = & -\f{1}{\sqrt{-\hat{g}}}\Big(\sqrt{-\hat{g}}\, \hat{g}^{c\hat{x}}\Big)_{,c} \;\;,
\end{eqnarray}
and from \eqref{eq:BS_conf}, \eqref{eq:def_sigma}, and $\widehat{\nabla}_a\widehat{\nabla}_b\ell = \hat{\Gamma}^{\hat{x}}_{ab}$  follows
\begin{subequations}\label{eq:Sigma_metric}
\begin{eqnarray}
\widehat{\Sigma}_{\hat{u}\hat{u}} & = & \hat{g}^{\hat{x}a}\hat{g}_{\hat{u}a,\hat{u}}
  -\f{1}{2}\hat{g}^{\hat{x}a}\hat{g}_{\hat{u}\hat{u},a}
  -\f{1}{4}g_{\hat{u}\hat{u}}\Theta\;\;, \\
\widehat{\Sigma}_{\hat{u}\hat{x}} & = & \hat{g}^{\hat{x}a }\hat{g}_{\hat{u}[\hat{x},a]}
  +\f{1}{2}\hat{g}^{\hat{x}\hat u}\hat{g}_{\hat u \hat x,\hat{u}}
  -\f{1}{4}\hat{g}_{\hat{u}\hat{x}}\Theta \\
\widehat{\Sigma}_{\hat{u}A} & = & \hat{g}^{\hat{x}b}\hat{g}_{\hat{u}[b,A]}
  +\f{1}{2}\Big(\hat{g}^{\hat{x}\hat{u}}\hat{g}_{\hat{u}A,\hat{u}} +\hat{g}^{\hat{x}B}\hat{g}_{AB,\hat{u}}\Big)\nonumber\\
  &&
  -\f{1}{4}\hat{g}_{\hat{u}A}\Theta\;\;  \\
  \widehat{\Sigma}_{\hat{x}\hat{x}} & = & \hat{g}^{\hat{u}\hat{x}}\hat{g}_{\hat{u}\hat{x},\hat{x}}\;\; 
    \end{eqnarray}
  \begin{eqnarray}
\widehat{\Sigma}_{\hat{x}A} & = & \hat{g}^{\hat{x}\hat{u}}\hat{g}_{\hat{u}(\hat{x},A)}
  +\f{1}{2}\hat{g}^{\hat{x}B}\hat{g}_{AB,\hat{x}} \;\;, \\
\widehat{\Sigma}_{AB} & = & \hat{g}^{\hat{x}\hat{u}}\hat{g}_{\hat{u}(A,B)}+\hat{g}^{\hat{x}C}\hat{g}_{C(A,B)}
  -\f{1}{2}\hat{g}^{\hat{x}a}\hat{g}_{AB,a}\nonumber\\
&&
  -\f{1}{4}\hat{g}_{AB}\Theta \, 
\end{eqnarray}
\end{subequations}
The calculation of $\Psi_4$ in \cite{Babiuc2009} involves  three different metrics:
\begin{enumerate}
  \item the Bondi--Sachs metric $g_{ab}$ of the physical spacetime,
  \item the conformal metric $\widehat{g}_{ab}$ that maps null infinity to a finite value in the
  conformal space time,
  \item  and  the conformal inertial metric $\tilde{g}_{ab}$. 
\end{enumerate}
As the conformal metric $\widehat{g}_{ab}$ is related to the Bondi--Sachs metric with the conformal factor $\ell$, the conformal Bondi metric $\tilde{g}_{ab}$ is related to the Bondi--Sachs metric with another conformal factor $\Omega$ like $\tilde{g}_{ab}=\Omega^2g_{ab}$. 
As in \cite{News, Babiuc2009}, we set
\begin{equation}
\label{eq:defOmega}
\Omega(u,\,x,\,x^A):=\ell(x)\omega(u,\,x^A)\;\;.
\end{equation}
This choice 
for the  conformal factor $\Omega$ (together with the particular choice of the function $r(x)$ and $\ell(x)$) has the advantage that the compactification in radial direction is decoupled from the angular behavior of the metric at $\ms{I}$. The definition \eqref{eq:defOmega} relates the conformal metric $\hat{g}_{ab}$ and the conformal inertial metric $\tilde{g}_{ab}$ like $\tilde{g}_{ab} =\omega^2\hat{g}_{ab}$.

Suppose $\tilde{x}^a$ are inertial coordinates as in the previous section such that the metric $\tilde{g}_{ab}$ at $\ms{I}$ has the form as in \eqref{eq:BondiFrame}. In this coordinate system we choose two real  null vectors $\tilde{n}^a:=\tilde{g}^{ab}\widetilde{\nabla}_b\Omega|_\ms{I}$ and $\tilde{l}^a:=\tilde{g}^{ab}\widetilde{\nabla}_b\tilde{u}|_\ms{I}$ and a complex null vector $\tilde{Q}^a$ that obey  $\tilde{l}^a\tilde{n}_a=-1$, $\tilde{Q}^a\overline{\tilde{Q}}_a=q$  , whereas all other scalar product vanish between them. 
These null vectors 
define a null tetrad $\tilde{z}^a_{(b)}:=(\tilde{l}^a,\,\tilde{n}^a,\,\tilde{Q}^a,\,\overline{\tilde{Q}^a})$  allowing us to write  the inertial conformal metric at $\ms{I}$ as  
  \begin{equation}
  \label{eq:in_con_metric_tetrad}
      \tilde{g}_{ab} = -\tilde{l}_a\tilde{n}_b - \tilde{n}_a\tilde{l}_b + \f{1}{q}\Big(\tilde{Q}_a\overline{\tilde{Q}}_b  + \overline{\tilde{Q}}_a\tilde{Q}_b\Big)
   \end{equation}
With this null tetrad, the Weyl scalar $\Psi_4$ is given by a contraction of the Weyl tensor \cite{W1988, Babiuc2009}  
\begin{equation}
\label{eq:Weyl_inertial}
\Psi_4 =-\f{1}{q} \lim_{\Omega \rightarrow 0} \bigg(\f{\tilde{n}^a\tilde{Q}^b\tilde{n}^c\tilde{Q}^d\tilde{\mc{C}}_{abcd}}{\Omega}\bigg)\;,
\end{equation}
which  corresponds to $-(1/q)\overline{\Psi}^{(P)}_4$   in the standard Newman-Penrose notation \cite{NP1962}, where $\Psi^{(P)}_4$  is the Weyl scalar as defined in \cite{Penrose1963}. 
   
The above defined null tetrad is not completely fixed \cite{NU1962, JN1965} as one still  has still the following three freedoms in the tetrad representation of the metric \eqref{eq:in_con_metric_tetrad}: (i) Lorentz transformations with the boost factor $\alpha$ and spatial rotations around the angle $\vartheta$ (where $\alpha$ and $\vartheta$ are real functions),  \ie
\begin{subequations}
 \begin{eqnarray}
l^{a^\prime} & = & \alpha \tilde{l}^a\;,\quad
n^{a^\prime}  =\f{1}{\alpha} \tilde{ n}^a \;,\quad
Q^{a^\prime}= e^{i\vartheta}\tilde{Q}^a \;, 
\end{eqnarray}
\end{subequations}
(ii) null rotations around $n^a$ with a complex function  $\kappa$
\begin{subequations}
\begin{eqnarray}
l^{a^\prime} & = & \tilde{l}^a+\overline{\kappa}\tilde{Q}^a + \kappa \overline{\tilde{Q}}^a  + \kappa\overline{\kappa}\tilde{n}^a\,,\;
n^{a^\prime} = \tilde{ n}^a\;, \\
Q^{a^\prime}&=& \tilde{Q}^a + \lambda \tilde{n}^a\;,
\end{eqnarray}
\end{subequations}
and (iii) null rotations around $l^a$ with a complex function $\lambda$
\begin{subequations}
\begin{eqnarray}
l^{a^\prime} & = & \tilde{l}^a \;,\;
n^{a^\prime} =  \tilde{n}^a + \overline{\lambda}\tilde{O}^a + \lambda \overline{\tilde{O}}^a  + \lambda\overline{\lambda}\tilde{l}^a\;,\\
Q^{a^\prime}&=& \tilde{Q}^a + \lambda\tilde{ l}^a\;.\label{eq:null_rot_Q}
\end{eqnarray}
\end{subequations}
Since the Weyl scalar \eqref{eq:Weyl_inertial} is invariant under null rotation around $\tilde{n}^a$ on $\ms{I}$, null rotations around $\tilde n^a$ can be used to calculate $\Psi_4$ in another frame at $\ms{I}$.

Babiuc \etal \cite{Babiuc2009} use four ingredients to find the relation between $\Psi_4$ in an inertial conformal frame and an arbitrary conformal frame on $\ms{I}$. The first ingredient is that the inertial conformal metric $\tilde{g}_{ab}$ and the conformal metric $\hat{g}_{ab}$ are related via $\tilde{g}_{ab}=\hat{\omega}^2\hat{g}_{ab}$. This implies that the Weyl tensor transforms between the  both frames  as \cite{Kuehnel2003}
\begin{equation}\label{eq:traf_weyl}
\tilde{C}_{abcd}= \hat{\omega}^2 \widehat{C}_{abcd}.
\end{equation}
At $\ms{I}$, the conformal 2-metric $\hat{g}_{AB}$ is subject to the constraint $\hat{g}_{AB} = (1/\hat{\omega}^2)q_{AB}$, which yields  the following elliptic equation \cite{News}
\begin{equation}\label{eq:det_omega}
\widehat{\mc{R}}\Big|_\ms{I}(\hat{g}_{AB}) = 2\Big(\hat{\omega}^2 +\hat{g}^{AB} \widehat{\nabla}_A\widehat{\nabla}_B\log\hat{ \omega}\Big)\Big|_{\ms{I}}\;\;,
\end{equation}
allowing one to calculate the conformal factor $\omega$ from the curvature scalar $\widehat{\mc{R}}$ of the surfaces $\hat{u}=const$ on $\ms{I}$. 
The second ingredient concerns the expression 
of the null vector $\tilde{n}^a$ in the conformal frame at  $\ms{I}$, \ie
\begin{equation}
\label{eq:conf_fraf_n}
 \tilde{n}^a |_\ms{I} =  \f{1}{\hat{\omega}}\hat{n}^a|_\ms{I}\;\;,
\end{equation}
where $\hat{n}^a$ is given by
\begin{equation}
\label{eq:n_conf}
\hat{n}^a\big|_\ms{I} = \hat{g}^{ab}\widehat{\nabla}_b \ell\big|_\ms{I} = -\hat{g}^{a\hat{x}}\big|_\ms{I} =\Big(-\hat{g}^{\hat{u}\hat{x}}|_\ms{I},\,0,\,-\hat{g}^{\hat{u}\hat{A}}|_\ms{I}\Big).
\end{equation}
Equation \eqref{eq:conf_fraf_n} shows that  the transformation between the inertial conformal frame and an arbitrary conformal frame corresponds to a Lorentz transformation (boost) from one frame to the other. The third ingredient is to use 
\eqref{eq:null_rot_Q} to transform $\tilde{Q}^a$ to an arbitrary conformal frame \footnote{Ref. \cite{News} and Ref. \cite{Babiuc2009}, include for numerical purposes a spatial rotation which is not required here.}, \ie
\begin{equation}
\label{eq:Q_inert_as_F}
\tilde{Q}^a \Big|_\ms{I} = \f{1}{\hat{\omega}}\hat{M}^a\Big|_\ms{I} + \f{\lambda}{\omega} \hat{n}^a\big|_\ms{I}\;\;, 
\end{equation}
where $\hat{M}^a|_{\ms{I}}:=(0,\,0,\,\hat{F}^A/R(a))$ such that $\hat{M}^a\overline{\hat{M}}_a = q$   and $\hat{M}^a{\hat{M}}_a = 0$ at $\ms{I}$.
Combining  \eqref{eq:Weyl_inertial}, \eqref{eq:traf_weyl}, \eqref{eq:conf_fraf_n}, and \eqref{eq:Q_inert_as_F}, gives the Weyl scalar $\Psi_4$ in an arbitrary conformal frame at $\ms{I}$  
 \begin{equation}
\label{eq:psi_4_conf}
\Psi_4\Big|_\ms{I} = -\f{1}{q}\f{1}{\hat{\omega}^3} \lim_{\hat{x}\rightarrow a} \Big(\f{\hat{n}^a\hat{M}^b\hat{n}^d\hat{M}^d\widehat{C}_{abcd}}{a-\hat{x}}\Big)\;,
\end{equation}
where the relation $\hat{\mc{C}}_{abcd}\hat n^a \hat M^b \hat n^v\hat n^d=0$ was used.
The main result of \cite{Babiuc2009} and the fourth ingredient for the determination of $\Psi_4$ is that \eqref{eq:psi_4_conf} can be expressed by the vector field $\widehat{\Sigma}_{ab}$ like  
\begin{equation}
\label{eq:Psi4_fin_full}
\Psi_4\Big|_\ms{I} = \f{1}{q}\f{1}{\hat{\omega}^3}\hat{n}^a\hat{M}^B\hat{M}^C\Big(\widehat{\nabla}_a \widehat{\Sigma}_{BC} - \widehat{\nabla}_B \widehat{\Sigma}_{aC}\Big)\Big|_\ms{J}\;\;.
\end{equation}
This equation has an advantage to \eqref{eq:psi_4_conf};  it is easier to determine from the metric at $\ms{I}$ than the (rather tedious) calculation of the contractions of the Weyl tensor.
\subsubsection{Calculation of $\Psi_4$ for SPIN-2}
We find the Weyl scalar $\Psi_4$ for SPIN-2 by deriving first the corresponding expression to \eqref{eq:Psi4_fin_full} in a quasi-spherical expansion while assuming the following limiting non-trivial values of $\hat{g}_{ab}^{(0)}$ and $\hat{g}_{ab}^{(1)}$ 
\begin{subequations}\label{eq:qs_metric}
\begin{eqnarray}
\hat{g}_{\hat{u}\hat{x}}^{(0)}|_\ms{I} & = &- R(a)\;,\; 
\hat{g}_{AB}^{(0)}|_\ms{I}  =  R^2(a)q_{AB}\;,  \\
\hat{g}_{\hat{u}\hat{x}}^{(1)}|_\ms{I} & = &- R(a)\;,\;
\hat{g}_{\hat{u}\hat{A}} ^{(1)}|_\ms{I}  =  -R^2(a){q}_{AB}\hat{U}^{B}_{(1)}|_\ms{I}\;, \;\\
\hat{g}_{\hat{A}\hat{B}} ^{(1)}|_\ms{I}  &=&  R^2(a)\hat{\gamma}^{(1)}_{AB} |_\ms{I}\;.
\end{eqnarray} 
\end{subequations}
To determine the conformal factor $\omega$ at $\ms{I}$, we consider  the quasi-spherical expansion 
\begin{equation}
\label{ }
\hat{\omega}(\varepsilon) \hat{=} \stackrel{0}{ \hat{\omega}}+ \stackrel{1}{\hat{\omega}}\varepsilon \;\;.
\end{equation}
Calculating  zero order terms in $\varepsilon$ of $\hat{g}_{AB}(\varepsilon) = q_{AB}/\hat{\omega}^2(\varepsilon)$ at $\ms{I}$ yields $\stackrel{0}{\hat{\omega}} = 1/R(a)$. To determine $\stackrel{1}{\hat{\omega}}$, we calculate the $O(\varepsilon)-$ contribution of \eqref{eq:det_omega} while using $\stackrel{0}{\hat{\omega}}=1/R(a)$ which gives us the equation 
\begin{eqnarray}
\f{1}{2R(a)}D^AD^B\hat{\gamma}^{(1)}_{AB}\Big|_\ms{I} & = & D^AD_A \stackrel{1}{\hat{\omega}}+ 2\stackrel{1}{\hat{\omega}}\;,
\end{eqnarray}
or at $\ms{I}$ in terms of the $\eth-$operator  
\begin{eqnarray}\label{eq:dgl_omega_1}
\f{1}{4qR(a)}\Big(\eth^2\overline{\hat{\mc{J}}}|_\ms{I}+\overline{\eth}^2\hat{\mc{J}}|_\ms{I}\Big) & = & \overline{\eth}\eth\stackrel{1}{\hat{\omega}}+ 2\stackrel{1}{\hat{\omega}}\;\;.
\end{eqnarray}
Since $\hat{\omega}_1$ is a real scalar field it has spin weight zero, therefore we assume for $\hat{\omega}_1$ an expansion in terms of $Z_{lm}$ like
\begin{equation}
\label{eq:omega_spin_dc}
\stackrel{1}{\hat{\omega}}\Big|_{\ms{I}}(\hat{x}^a) = \sum_{l=0}^\infty\sum_{m=-l}^m \hat{\omega}^{lm}(\hat{u})Z_{lm}(\hat{x}^A)\;\;.
\end{equation}
Inserting \eqref{eq:omega_spin_dc} and \eqref{eq:exp_J_I} into \eqref{eq:dgl_omega_1} while using \eqref{eq:prop_sYlm} and \eqref{eq:def_Zlm} implies the spectral coefficients $\hat{\omega}^{lm}$  
\begin{eqnarray}
\hat{\omega}^{00} & = & \hat{\omega}^{1(-1)} = \hat{\omega}^{10} =\hat{\omega}^{11}=0\;\;,\\
\hat{\omega}^{lm}(\hat{u}) & = & \f{C_{lm}}{2^{l+3}\hat{u}^{l+2}q}\f{\sqrt{l-1}}{[2-l(l+1)]\sqrt{l(l+1)(l+2)}} \nonumber\\
&& (l>1) \mbox{ and } |m|\le l\;\;.
\end{eqnarray}
The quasi-spherical approximation of  null vector $\hat{n}^a$  can be found using \eqref{eq:n_conf}, \ie 
\begin{eqnarray}
\hat{n}^a|_\ms{I} & \hat{=} & -\f{1}{R(a)} \Big(\kron{a}{\hat{u}} + \varepsilon \hat{U}^A_{(1)}\Big|_{\ms{I}}\kron{a}{A}\Big)\;.
\end{eqnarray}
To find $\hat M^A$ at $\ms{J}$, we use its quasi-spherical expansion 
\begin{equation}
\label{ }
\hat{M}^A\Big|_\ms{J} \hat= \f{1}{R(a)}\Big(\stackrel{0}{F}\!^A+\varepsilon  \stackrel{1}{F}\!\!\Big)\;,
\end{equation}
 its normalization $\hat{M}^a\hat{\overline M}_a-q=\hat M^a \hat M_a =0$, the quasi-spherical expansion for the metric \eqref{eq:qs_metric}, and  the relations \eqref{eq:qab_dyad}, and \eqref{eq:gamAB_dyad} which allow us to deduce
 \begin{equation}
\label{ }
\stackrel{0}{F}\!^A = q^A\;\;,\quad
\stackrel{0}{F}\!^A =- \f{\hat {\mc{J}}}{2q}\overline{q}^A\;\;.\quad
\end{equation}

An inspection of \eqref{eq:Sigma_metric} at $\ms{I}$ shows that $\Sigma_{ab}$ is $O(\varepsilon)$ at $\ms{I}$. Therefore, $\Psi_4$ is of $O(\varepsilon)$ at $\ms{I}$ and only the $O(\varepsilon^0)$ parts of $\omega,\,\hat{n}^a$, and $\hat{M}^A$ must be taken into account in its calculation. Since we have $n^{\hat{x}}|_\ms{I} = 0$ and because $\Sigma_{ab}$ is of  $O(\varepsilon)$ at $\ms{I}$, only the following covariant derivatives of $\Sigma_{ab}$ must be considered at $\ms{I}$
  \begin{eqnarray}
   \!\!\!\!\!\widehat\nabla_{\hat{u}}\hat{\Sigma}_{AB}\Big |_\ms{I} \!\!\!\!& \hat{=} &\!\!\!\!
     \bigg[q_{C(A}D_{B)}U^C_{(1),\hat{u}}+\f{1}{2}\gamma^{(1)}_{AB,\hat{u}\hat{u}}\bigg]R(a)\varepsilon \;,\\
      \widehat\nabla_{C}\hat{\Sigma}_{\hat{u}B}\Big|_\ms{I} \!\! \!\!& \hat{=} & \!0\;.
\end{eqnarray}
With these covariant derivatives we find  $\Psi_4$ in the quasi-spherical expansion as  
\begin{eqnarray}
\label{eq:psi_4_lin}
\!\!\!\!\hat{\Psi}_4\Big|_\ms{I} &\hat{=} & \bigg(\sqrt{\f{q}{2}}\eth \mc{U}_{,\hat{u}}+\f{1}{2} \mc{J}_{,\hat{u}\hat{u}}
   \bigg)\Bigg|_{\hat{x}=a}R({a}) \varepsilon\;\;,
\end{eqnarray}
which corrects equation (3.54)  in \cite{Babiuc2009}. 
Using \eqref{eq:exp_J_I}, \eqref{eq:exp_U_I}, and \eqref{eq:psi_4_lin} for SPIN-2 yields a simple expression for $\hat{\Psi}_4$ at null infinity

\begin{eqnarray}
\label{eq:psi_4_spin2}
\!\!\!\!\!\!\!\hat{\Psi}_4\Big|_{\ms{I}}\!\!\!\!\!\! &\hat{=}&\!\!\!\varepsilon R(a)\!\sum_{l=2}^\infty \sum_{m=-l}^l\bigg[\f{(l+3)C_{lm}}{2^{l} \hat{u}^{l+4}l(l+1)}\bigg]\,_2Y_{lm}(\hat{x}^A) 
\end{eqnarray}
\section{Summary}\label{sec:Discuss}
\noindent
We discussed a linearized vacuum solution of the Einstein equations in the Bondi--Sachs formulation of General Relativity. Assuming that the metric obeys regularity conditions along the central geodesic tracing the vertices of the null cones, we found that the spherically symmetric background spacetime is  Minkowskian. We then derived a differential equation for a two-tensor, $\psi_{AB}$,  whose solution allows one to determine  in a hierarchal manner the linear perturbations $\gamma_{AB}^{(1)}$, $U^A_{(1)}$, and $\Phi_{(1)}$. Utilizing  a representation of the unit sphere with a complex dyad,  it was shown that the differential equation for $\psi_{AB}$ is a wave equation for a spin-2 field $\psi$.The field $\psi$ is corresponds to  $2r\Psi_0$, where $\Psi_0$ is the Newman--Penrose Weyl scalar \cite{NP1962}. We reformulated the hierarchal equations for  $\gamma_{AB}^{(1)}$, $U^A_{(1)}$, and $\Phi_{(1)}$ as differential equations for spin weighted variables $\mc{J},\,\mc{U}$ and $\breve{\Phi}$, respectively. Since the function $\psi$ determines all linearized perturbations {\it in vacuo}, we refer to it as a {\it master function}.  Under the assumption of the existence of a  power series of $\psi$ in terms of the areal distance $r$ at $r=0$, we solved  the   equation for $\psi$ locally, and subsequently, those for  $\mc{J},\,\mc{U}$ and ${\Phi}_{(1)}$, at the vertices. This provided us (eq. \eqref{eq:reg_BS}) with the linearized boundary conditions  for  the Bondi--Sachs metric functions in a spin-representation  at the vertex in vacuum spacetimes. It also generalizes previously presented axially symmetric boundary conditions  \cite{MM2013} to the three-dimensional case with no symmetries. These boundary conditions may be used in  numerical simulations to calculate vacuum space times  in the Bondi--Sachs framework  when the vertex of the null cones is part of the numerical grid.  

We employed the boundary conditions for $\psi$ in solving  the wave equation for $\psi$ globally by two different approaches. In addition, we required the solution for $\psi$ to be finite at large distances to assure an asymptotically flat solution for the perturbations $J,\,U$, and ${\Phi}_{{(1)}}$. In both approaches, we represented $\psi$ as power series in terms of spin-2 spherical harmonics with coefficients depending on the retarded time $u$ and the radius $r$. This allowed us to deduce a partial differential equation in terms of $u$ and $r$ for the coefficients of this series. 

In the first approach, we imposed a standard ansatz of separation of variables to solve the differential equation for the coefficients of the series. In the procedure we determined an ordinary inhomogeneous differential equation (eq. \eqref{eq:ode_Az}) that is  most generally solved by a finite spectral series using polynomial coefficients with modified spherical Bessel functions of the first and second kind as base functions. As the modified spherical Bessel functions of second kind are singular at the origin, they must be discarded by the regularity conditions at the origin, whereas a solution purely depending on the modified spherical Bessel functions of first kind obeys this regularity condition.  Although, this solution (eq. \eqref{eq:psi_bessel}) for the coefficients of the spectral series of $\psi$ is regular at the origin, we discarded it, because it diverges exponentially as $r$ tends to infinity, and it would have generated a solution for the perturbations that is not asymptotically flat. Hence, regularity of the Bondi--Sachs metric  at the vertex is not a sufficient requirement to obtain an asymptotically flat solution of the Einstein equation in the Bondi--Sachs framework. Our calculation also demonstrated that using a standard separation of variables, where the function is decomposed into a product of four factors of which each depends on one of the coordinate $x^a$ only,  is unsuited to solve the wave equation in Bondi--Sachs coordinates, if one requires the solution to be regular at the vertex and asymptotically finite.

In the second approach, we transformed the second-order wave equation for $\psi$ into a inhomogeneous first-order transport equation (eq. \eqref{eq:idt_w_lm}). The inhomogenity of this equation vanishes for the lowest ($l=2$) spin-2 harmonic. The corresponding homogeneous transport equation has the characteristic surface $u+2r=const$. Since the wave equation for $\psi$ and the transport equation are related by a linear integral transformation,  the solution of the wave equation for $\psi$ and the corresponding transport equation have the same characteristic for the lowest multipole.  Using a polynomial ansatz (eq. \eqref{eq:ans_char}) that incorporates the characteristic information of the wave equation, allowed us  to find a solution for {\it all} multipoles for the master function and linearized perturbations (eq. \eqref{eq:sol_SPIN2}) that are  regular at the origin and asymptotically flat. 
This solution is referred to as SPIN-2, because it represents spin-2 waves propagating on a Minkowski background spacetime. SPIN-2 has some advantages in regard to other linearized solutions in the Bondi--Sachs framework found in the literature. First, SPIN-2 is given by  simple rational expressions in terms of the spin-weighted  quantities of the Bondi--Sachs metric.  Second, it describes all multipole of the perturbations, while other solutions give  only their lowest multipoles  \cite{B2005, BK2009} or require an elaborated procedure to generate those  multipoles \cite{Gomez1994, CCE1996}. 
    
For the SPIN-2 solution, we calculated the Weyl scalar $\Psi_4$ at null infinity  using the formalism of \cite{Babiuc2009}.   For pedagogical reasons, we also   summarized the formalism of \cite{Babiuc2009} by pointing out the four most important steps in obtaining a simpler formula for $\Psi_4$ at null infinity in linearized gravity. This simple analytical expression (eq. \eqref{eq:psi_4_spin2}) for $\Psi_4$ at null infinity and the explicit form of the perturbations $\mc{J},\,\mc{U},$ and ${\Phi}_{(1)}$ (eq. \eqref{eq:sol_SPIN2}) make SPIN-2 an ideally suited testbed solution for simulations in the Bondi--Sachs framework and to test numerical wave extraction methods at null infinity, \eg \cite{RPB2012} describes the most recent process of such simulations (for others see \cite{WinicourLRR}). 
      
In the future, we plan to test the stability of the  SPIN-2 solution against small perturbations, to investigate its physical reliability. It also would be  interesting to utilize SPIN-2 to study quadratic perturbations with respect to a Minkowskian background spacetime. 
\section{Acknowledgment}
It is a pleasure to thank E. M\"uller, J. Winicour, A. J. Penner, L. Lehner, P. Jofr\'e Pfeil for comments on the manuscript and discussions. I also thank B. Carter for providing me the proceedings of ref. \cite{Friedrich1988} and \cite{W1988}. Financial support is appreciated from the Max Planck Society, from the Collaborative Research Center on Gravitational Wave Astronomy of the Deutsche Forschungsgemeinschaft (DFG SFB/Transregio 7), the Observatory of Paris, and the CNRS. 
\vspace{1ex}

\begin{appendix}

\section{Ricci tensor contributions for the main equations}\label{app:Ricci}

\subsection{Contributions at $O(\varepsilon^0)$}\label{app:O_0}
The non-vanishing Ricci tensor contributions of the  hypersurface and evolution equations are at $O(\varepsilon^0)$
\begin{eqnarray}
  \stackrel{0}{\mc{R}}_{rr} & = & \f{1}{r}\f{d\beta_0}{dr}\label{eq:beta_back} \\
 \stackrel{0}{\mc{R}}_{(2D)} & = & 2 e^{-2\beta_0}\f{d}{dr}\left[r\Big(1+2\Phi_0+2\beta_0\Big)\right] -\ms{R}(q_{AB})\nonumber\\
 &&\label{eq:Phi_back} 
\end{eqnarray}
where $\ms{R}(q_{AB})=2$ is the Ricci curvature scalar of the unit sphere. 
\subsection{Contributions at $O(\varepsilon^1)$}
\noindent
The relevant Ricci tensor contributions for the hypersurface  equations at $O(\varepsilon)$ are  
\begin{eqnarray}
 \stackrel{1}{\mc{R}}_{rr} & = & \f{1}{r}\Big[\beta_{(1)}\Big]_{,r}\;\;, \label{eq:beta_lin} 
 \end{eqnarray}
\begin{eqnarray}
  \stackrel{1}{\mc{R}}_{rA} & = & \f{1}{2r^2}\left[r^4e^{-2\beta_0}q_{AE}U^E_{(1),r}\right]_{,r} 
   -r^2\left[\f{1}{r^2}D_A\beta_{(1)}\right]_{,r}\;\;,
   \label{eq:UA_lin} \nonumber \\
&&+\f{1}{2} q^{EF}D_E \gamma^{(1)}_{AF,r}
\end{eqnarray}
\begin{eqnarray}
 \stackrel{1}{\mc{R}}\!^{(2D)} & = &
      4e^{-2\beta_0}\Big[re^{2\Phi_0+2\beta_0}\Phi_{(1)}\Big]_{,r}
      -
D^A D^B \gamma^{(1)}_{AB}\nonumber\\
 &&
  + 2q^{AB}D_A D_B \beta_{(1)}
  -\f{1}{r^2}e^{-2\beta_0}D_A\Big[r^4U^A_{(1)}\Big]_{,r}\;\;,\label{eq:Phi_lin}
\end{eqnarray}  
 and those for the evolution equations
\begin{eqnarray}
 \stackrel{1}{\mc{R}} \!^{(TT)}_{AB} &=&
    r(r\gamma^{(1)}_{AB,u})_{,r}-\f{1}{2}\left(r^2e^{2\Phi_0+2\beta_0}\gamma^{(1)}_{AB,r}\right)_{,r}
    \nonumber\\
   &&   -2e^{2\beta_0} D_AD_B\beta_{(1)}
     + q_{AE}D_B\Big(r^2U^E_{(1)}\Big)_{,r}\nonumber\\
    &&
     -\f{1}{2} q_{AB}D_E\Big[r^2U^E_{(1)}\Big]_{,r} \;\;.     \label{eq:R_TT}
\end{eqnarray}
\section{The master equation and the flat-space scalar wave equation}\label{secAPPB}
In this appendix, it is shown how the master equation relates to a flat space wave equation of scalar field with spin weight zero. This offers a comparison to the approach of \cite{CCE1996}, where the perturbations are generated by a spin-0 fields.

The homogeneous flat space  wave equation for a spin-0 field $h$  is  
\begin{equation}
\label{WES0}
0=\square h,
\end{equation}
where  $\square h:= \eta^{ab}\nabla_a\nabla_b h$ is the d'Alembert operator, which reads in outgoing polar null coordinates 
\begin{equation}
\label{ }
r^2\square f =2r(rh)_{,ur} -(r^2h_{,r})_{,r}-\overline \eth\eth h .
\end{equation}

If we commute the eth and eth bar operator in the master equation\eqref{eq:dgl_psi}, we obtain
\begin{equation}
\label{eqdgl_psi_commeth}
0=\f{1}{r^2}\Big[r^4(2\psi_{,u}-\psi_{,r})\Big]_{,r} -\eth\overline \eth \psi\;.
\end{equation}
Setting $\psi:=\eth^2{F}$, where $F$ has the spin weight zero, and inserting this new definition into \eqref{eqdgl_psi_commeth} yields 
\begin{equation}
\label{eqdgl_F_s2}
0=\eth^2\bigg\{\f{1}{r^2}\Big[r^4(2{F}_{,u}-{F}_{,r})\Big]_{,r} -(\overline \eth\eth -2)\ {F}\bigg\}\;,
\end{equation}
where we again commuted the eth and ethbar operators to factor out $\eth^2$. Eq.  \eqref{eqdgl_F_s2} implies for ${F}$ the following differential equation
\begin{equation}
\label{eqdgl_F_s2}
0=\f{1}{r^2}\Big[r^4(2{F}_{,u}-{F}_{,r})\Big]_{,r} -(\overline \eth\eth +2)\ {F}\;.
\end{equation}
To find how \eqref{eqdgl_F_s2} relates to \eqref{WES0} we introduce a further spin-0 field $F=r^n f$ and insert this definition into \eqref{eqdgl_F_s2} which yields after dividing out $r^n$
\begin{eqnarray}
\label{eqdgl_f}
0&=&2r(rf_{,r})_{,u}-(r^2f_{,r})_{,r}-\overline \eth\eth f +2r(n+3)f_{,u}\nonumber\\
&&-2r(n+1)f_{,r}-(n+2)(n+1) f.
\end{eqnarray}
It can be seen that the first three terms in \eqref{eqdgl_f} correspond to $r^2\square f$ which is the principle part of the flat space scalar wave equation in outgoing Bondi--Sachs coordinates. The other additional  terms are  indicate that \eqref{eqdgl_f} is a a quasispherical wave equation. In fact, it is not possible to set all these terms to zero for any number of $n$. The linear part in  \eqref{eqdgl_f} vanishes if $n=-1$ or $n=-2$, i.e. in this case there are no restoring forces. In particular, if $n=-1$ and if $f$ is time-independent then \eqref{eqdgl_f} is the standard Laplace equation in spherical coordinates. Whereas if $f$ is time-dependent, we have the wave equation
\begin{equation}
\label{eq:we_damp}
r^2\square f =- 4 r f_{,u}\;\;,
\end{equation}
with an  additional damping term $-4rf_{,u}$. Equation \eqref{eq:we_damp} has also been obtained by \cite{W2013} in a different approach. For $n=-2$,  eq. \eqref{eqdgl_f} becomes
\begin{equation}
\label{eq:we_notgood}
r^2\square f =- 2 r( f_{,u}+f_{,r})\;\;,
\end{equation}
which is a wave equation with a damping and a force term.  This equation has the disadvantage to \eqref{eq:we_damp} that it does not reduce to the Laplace equation for time-independent fields. Therefore \eqref{eq:we_damp} is preferable to \eqref{eq:we_notgood}, and we conclude that the master function $\psi$ is related to the spin-0 field $f$ via \eqref{eq:we_damp} and 
\begin{equation}
\label{def_psi_f}
\psi=\f{1}{r}\eth^2 f\;\;. 
\end{equation}

Based on \eqref{eq:we_damp} and \eqref{def_psi_f}, we now sketch an alternative spin-0 approach to the one given in \cite{CCE1996}. Let $\alpha$ and $\mc{Z}$ be spin-0 field that are related to the perturbations $\mc{J} $ and $\mc{Z}$ via
\begin{subequations}\label{def_JU}
\begin{eqnarray}
\mc{J} &: = & \eth^2\alpha \\
\mc{U} &: = &\f{1}{\sqrt{2q}} \eth\mc{Z}
\end{eqnarray} 
\end{subequations}
Inserting \eqref{def_JU} into \eqref{eq:dgl_psi_gam}-\eqref{eq:EE_1_Phi_eth} while using \eqref{def_psi_f} allows us to deduce three equations
\begin{subequations}\label{spin0_pert}
\begin{eqnarray}
0&=& (r\alpha)_{,rr} - \f{f}{r}\;\;,\label{alphaeqn}\\
0&=&  \Big(r^4\mc{Z}_{,r}\Big)_{,rr}  + (\bar\eth\eth+2)f\label{zeqn}\\
0&=&  \widetilde\Phi_{,r}
     -\f{1}{q} \, \bar\eth\eth\bigg[(\bar{ \eth}\eth+2)  \alpha
  +\f{2}{r^2}\Big(r^4\mc{Z}\Big)_{,r}\bigg],\label{Phieqn}
\end{eqnarray}
\end{subequations}
which can be used to the determine the perturbations $\mc{J},\,\mc{U}$, and $\widetilde \Phi$ in the following procedure:
(1) solve the damped wave equation \eqref{eq:we_damp}; (2) integrate eqs. \eqref{spin0_pert} according to the given hierarchy to find $\widetilde\Phi$;  and (3) use \eqref{def_JU} to obtain $\mc{J}$ and $\mc{U}$.

\end{appendix}
\bibliographystyle{apsrev}
\bibliography{LinWavePRD}{}

\begin{thebibliography}{38}
\expandafter\ifx\csname natexlab\endcsname\relax\def\natexlab#1{#1}\fi
\expandafter\ifx\csname bibnamefont\endcsname\relax
  \def\bibnamefont#1{#1}\fi
\expandafter\ifx\csname bibfnamefont\endcsname\relax
  \def\bibfnamefont#1{#1}\fi
\expandafter\ifx\csname citenamefont\endcsname\relax
  \def\citenamefont#1{#1}\fi
\expandafter\ifx\csname url\endcsname\relax
  \def\url#1{\texttt{#1}}\fi
\expandafter\ifx\csname urlprefix\endcsname\relax\def\urlprefix{URL }\fi
\providecommand{\bibinfo}[2]{#2}
\providecommand{\eprint}[2][]{\url{#2}}

\bibitem[{\citenamefont{{Bondi} et~al.}(1962)\citenamefont{{Bondi}, {van der
  Burg}, and {Metzner}}}]{Bondietal1962}
\bibinfo{author}{\bibfnamefont{H.}~\bibnamefont{{Bondi}}},
  \bibinfo{author}{\bibfnamefont{M.~G.~J.} \bibnamefont{{van der Burg}}},
  \bibnamefont{and} \bibinfo{author}{\bibfnamefont{A.~W.~K.}
  \bibnamefont{{Metzner}}}, \bibinfo{journal}{\ppsla}
  \textbf{\bibinfo{volume}{269}}, \bibinfo{pages}{21} (\bibinfo{year}{1962}).

\bibitem[{\citenamefont{{Sachs}}(1962)}]{Sachs1962}
\bibinfo{author}{\bibfnamefont{R.~K.} \bibnamefont{{Sachs}}},
  \bibinfo{journal}{\ppsla} \textbf{\bibinfo{volume}{270}},
  \bibinfo{pages}{103} (\bibinfo{year}{1962}).

\bibitem[{\citenamefont{{Tamburino} and {Winicour}}(1966)}]{TW}
\bibinfo{author}{\bibfnamefont{L.~A.} \bibnamefont{{Tamburino}}}
  \bibnamefont{and} \bibinfo{author}{\bibfnamefont{J.~H.}
  \bibnamefont{{Winicour}}}, \bibinfo{journal}{\pr}
  \textbf{\bibinfo{volume}{150}}, \bibinfo{pages}{1039} (\bibinfo{year}{1966}).

\bibitem[{\citenamefont{{Winicour}}(2012)}]{WinicourLRR}
\bibinfo{author}{\bibfnamefont{J.}~\bibnamefont{{Winicour}}},
  \bibinfo{journal}{\lrr} \textbf{\bibinfo{volume}{15}}, \bibinfo{pages}{2}
  (\bibinfo{year}{2012}).

\bibitem[{\citenamefont{{Teukolsky}}(1982)}]{Teuk}
\bibinfo{author}{\bibfnamefont{S.~A.} \bibnamefont{{Teukolsky}}},
  \bibinfo{journal}{\prd} \textbf{\bibinfo{volume}{26}}, \bibinfo{pages}{745}
  (\bibinfo{year}{1982}).

\bibitem[{\citenamefont{{Rinne}}(2009)}]{Rinne}
\bibinfo{author}{\bibfnamefont{O.}~\bibnamefont{{Rinne}}},
  \bibinfo{journal}{\cqg} \textbf{\bibinfo{volume}{26}},
  \bibinfo{pages}{048003} (\bibinfo{year}{2009}).

\bibitem[{\citenamefont{{Winicour}}(1983)}]{W1983}
\bibinfo{author}{\bibfnamefont{J.}~\bibnamefont{{Winicour}}},
  \bibinfo{journal}{\jmp} \textbf{\bibinfo{volume}{24}}, \bibinfo{pages}{1193}
  (\bibinfo{year}{1983}).

\bibitem[{\citenamefont{{Winicour}}(1984)}]{W1984}
\bibinfo{author}{\bibfnamefont{J.}~\bibnamefont{{Winicour}}},
  \bibinfo{journal}{\jmp} \textbf{\bibinfo{volume}{25}}, \bibinfo{pages}{2506}
  (\bibinfo{year}{1984}).

\bibitem[{\citenamefont{{Winicour}}(1987)}]{W1987}
\bibinfo{author}{\bibfnamefont{J.}~\bibnamefont{{Winicour}}},
  \bibinfo{journal}{\grg} \textbf{\bibinfo{volume}{19}}, \bibinfo{pages}{281}
  (\bibinfo{year}{1987}).

\bibitem[{\citenamefont{{Newman} and {Penrose}}(1966)}]{N2BMS1966}
\bibinfo{author}{\bibfnamefont{E.~T.} \bibnamefont{{Newman}}} \bibnamefont{and}
  \bibinfo{author}{\bibfnamefont{R.}~\bibnamefont{{Penrose}}},
  \bibinfo{journal}{\jmp} \textbf{\bibinfo{volume}{\textbf{7}}},
  \bibinfo{pages}{863} (\bibinfo{year}{1966}).

\bibitem[{\citenamefont{{Papadopoulos}}(1994)}]{PapPhD}
\bibinfo{author}{\bibfnamefont{P.~O.} \bibnamefont{{Papadopoulos}}}, Ph.D.
  thesis, \bibinfo{school}{University of Pittsburgh.} (\bibinfo{year}{1994}).

\bibitem[{\citenamefont{{G{\'o}mez} et~al.}(1994)\citenamefont{{G{\'o}mez},
  {Papadopoulos}, and {Winicour}}}]{Gomez1994}
\bibinfo{author}{\bibfnamefont{R.}~\bibnamefont{{G{\'o}mez}}},
  \bibinfo{author}{\bibfnamefont{P.}~\bibnamefont{{Papadopoulos}}},
  \bibnamefont{and}
  \bibinfo{author}{\bibfnamefont{J.}~\bibnamefont{{Winicour}}},
  \bibinfo{journal}{\jmp} \textbf{\bibinfo{volume}{35}}, \bibinfo{pages}{4184}
  (\bibinfo{year}{1994}), \eprint{arXiv:gr-qc/0006081}.

\bibitem[{\citenamefont{{Lehner}}(1998)}]{LuisPhD}
\bibinfo{author}{\bibfnamefont{L.}~\bibnamefont{{Lehner}}}, Ph.D. thesis,
  \bibinfo{school}{University of Pittsburgh} (\bibinfo{year}{1998}).

\bibitem[{\citenamefont{{Bishop} et~al.}(1996)\citenamefont{{Bishop},
  {G{\'o}mez}, {Lehner}, and {Winicour}}}]{CCE1996}
\bibinfo{author}{\bibfnamefont{N.~T.} \bibnamefont{{Bishop}}},
  \bibinfo{author}{\bibfnamefont{R.}~\bibnamefont{{G{\'o}mez}}},
  \bibinfo{author}{\bibfnamefont{L.}~\bibnamefont{{Lehner}}}, \bibnamefont{and}
  \bibinfo{author}{\bibfnamefont{J.}~\bibnamefont{{Winicour}}},
  \bibinfo{journal}{\prd} \textbf{\bibinfo{volume}{54}}, \bibinfo{pages}{6153}
  (\bibinfo{year}{1996}).

\bibitem[{\citenamefont{{Goldberg} et~al.}(1967)\citenamefont{{Goldberg},
  {Macfarlane}, {Newman}, {Rohrlich}, and {Sudarshan}}}]{eth}
\bibinfo{author}{\bibfnamefont{J.~N.} \bibnamefont{{Goldberg}}},
  \bibinfo{author}{\bibfnamefont{A.~J.} \bibnamefont{{Macfarlane}}},
  \bibinfo{author}{\bibfnamefont{E.~T.} \bibnamefont{{Newman}}},
  \bibinfo{author}{\bibfnamefont{F.}~\bibnamefont{{Rohrlich}}},
  \bibnamefont{and} \bibinfo{author}{\bibfnamefont{E.~C.~G.}
  \bibnamefont{{Sudarshan}}}, \bibinfo{journal}{\jmp}
  \textbf{\bibinfo{volume}{8}}, \bibinfo{pages}{2155} (\bibinfo{year}{1967}).

\bibitem[{\citenamefont{{Glass} and {Naber}}(1997)}]{GN1997}
\bibinfo{author}{\bibfnamefont{E.~N.} \bibnamefont{{Glass}}} \bibnamefont{and}
  \bibinfo{author}{\bibfnamefont{M.~G.} \bibnamefont{{Naber}}},
  \bibinfo{journal}{\cqg} \textbf{\bibinfo{volume}{14}}, \bibinfo{pages}{1899}
  (\bibinfo{year}{1997}), \eprint{arXiv:gr-qc/9712073}.

\bibitem[{\citenamefont{{Newman} and {Penrose}}(1962)}]{NP1962}
\bibinfo{author}{\bibfnamefont{E.}~\bibnamefont{{Newman}}} \bibnamefont{and}
  \bibinfo{author}{\bibfnamefont{R.}~\bibnamefont{{Penrose}}},
  \bibinfo{journal}{\jmp} \textbf{\bibinfo{volume}{3}}, \bibinfo{pages}{566}
  (\bibinfo{year}{1962}).

\bibitem[{\citenamefont{{Bishop}}(2005)}]{B2005}
\bibinfo{author}{\bibfnamefont{N.~T.} \bibnamefont{{Bishop}}},
  \bibinfo{journal}{\cqg} \textbf{\bibinfo{volume}{22}}, \bibinfo{pages}{2393}
  (\bibinfo{year}{2005}), \eprint{arXiv:gr-qc/0412006}.

\bibitem[{\citenamefont{{Bishop} and {Kubeka}}(2009)}]{BK2009}
\bibinfo{author}{\bibfnamefont{N.~T.} \bibnamefont{{Bishop}}} \bibnamefont{and}
  \bibinfo{author}{\bibfnamefont{A.~S.} \bibnamefont{{Kubeka}}},
  \bibinfo{journal}{\prd} \textbf{\bibinfo{volume}{80}}, \bibinfo{eid}{064011}
  (\bibinfo{year}{2009}), \eprint{0907.1882}.

\bibitem[{\citenamefont{{Reisswig} et~al.}(2012)\citenamefont{{Reisswig},
  {Bishop}, and {Pollney}}}]{RPB2012}
\bibinfo{author}{\bibfnamefont{C.}~\bibnamefont{{Reisswig}}},
  \bibinfo{author}{\bibfnamefont{N.~T.} \bibnamefont{{Bishop}}},
  \bibnamefont{and}
  \bibinfo{author}{\bibfnamefont{D.}~\bibnamefont{{Pollney}}},
  \bibinfo{journal}{ArXiv/gr-qc:1208.3891}  (\bibinfo{year}{2012}),
  \eprint{1208.3891}.

\bibitem[{\citenamefont{{Manasse} and {Misner}}(1963)}]{MM1963}
\bibinfo{author}{\bibfnamefont{F.~K.} \bibnamefont{{Manasse}}}
  \bibnamefont{and} \bibinfo{author}{\bibfnamefont{C.~W.}
  \bibnamefont{{Misner}}}, \bibinfo{journal}{\jmp}
  \textbf{\bibinfo{volume}{4}}, \bibinfo{pages}{735} (\bibinfo{year}{1963}).

\bibitem[{\citenamefont{{M\"adler} and {M\"uller}}(2013)}]{MM2013}
\bibinfo{author}{\bibfnamefont{T.}~\bibnamefont{{M\"adler}}} \bibnamefont{and}
  \bibinfo{author}{\bibfnamefont{E.}~\bibnamefont{{M\"uller}}},
  \bibinfo{journal}{\cqg} \textbf{\bibinfo{volume}{30}},
  \bibinfo{pages}{055019} (\bibinfo{year}{2013}).

\bibitem[{\citenamefont{{Penrose}}(1963)}]{Penrose1963}
\bibinfo{author}{\bibfnamefont{R.}~\bibnamefont{{Penrose}}},
  \bibinfo{journal}{\prl} \textbf{\bibinfo{volume}{10}}, \bibinfo{pages}{66}
  (\bibinfo{year}{1963}).

\bibitem[{\citenamefont{{Babiuc} et~al.}(2009)\citenamefont{{Babiuc}, {Bishop},
  {Szil{\'a}gyi}, and {Winicour}}}]{Babiuc2009}
\bibinfo{author}{\bibfnamefont{M.~C.} \bibnamefont{{Babiuc}}},
  \bibinfo{author}{\bibfnamefont{N.~T.} \bibnamefont{{Bishop}}},
  \bibinfo{author}{\bibfnamefont{B.}~\bibnamefont{{Szil{\'a}gyi}}},
  \bibnamefont{and}
  \bibinfo{author}{\bibfnamefont{J.}~\bibnamefont{{Winicour}}},
  \bibinfo{journal}{\prd} \textbf{\bibinfo{volume}{79}}, \bibinfo{eid}{084011}
  (\bibinfo{year}{2009}), \eprint{0808.0861}.

\bibitem[{\citenamefont{{Misner} et~al.}((1973))\citenamefont{{Misner},
  {Thorne}, and {Wheeler}}}]{MTW}
\bibinfo{author}{\bibfnamefont{C.~W.} \bibnamefont{{Misner}}},
  \bibinfo{author}{\bibfnamefont{K.~S.} \bibnamefont{{Thorne}}},
  \bibnamefont{and} \bibinfo{author}{\bibfnamefont{J.~A.}
  \bibnamefont{{Wheeler}}}, \emph{\bibinfo{title}{Gravitation}}
  (\bibinfo{publisher}{San Francisco: W.H.~Freeman and Co.},
  \bibinfo{year}{(1973)}).

\bibitem[{\citenamefont{{Stewart}}(1993)}]{StewardBook}
\bibinfo{author}{\bibfnamefont{J.}~\bibnamefont{{Stewart}}},
  \emph{\bibinfo{title}{Advanced General Relativity}}
  (\bibinfo{publisher}{Cambridge, UK: Cambridge University Press},
  \bibinfo{year}{1993}), ISBN \bibinfo{isbn}{0521449464}.

\bibitem[{\citenamefont{{Penrose} and {Rindler}}(1987)}]{PenRind}
\bibinfo{author}{\bibfnamefont{R.}~\bibnamefont{{Penrose}}} \bibnamefont{and}
  \bibinfo{author}{\bibfnamefont{W.}~\bibnamefont{{Rindler}}},
  \emph{\bibinfo{title}{{Spinors and space-time. Vol. 1: Two-spinor calculus
  and relativistic fields.}}} (\bibinfo{publisher}{Cambridge, UK: Cambridge
  University Press}, \bibinfo{year}{1987}).

\bibitem[{\citenamefont{{G{\'o}mez} et~al.}(1997)\citenamefont{{G{\'o}mez},
  {Lehner}, {Papadopoulos}, and {Winicour}}}]{ethNR}
\bibinfo{author}{\bibfnamefont{R.}~\bibnamefont{{G{\'o}mez}}},
  \bibinfo{author}{\bibfnamefont{L.}~\bibnamefont{{Lehner}}},
  \bibinfo{author}{\bibfnamefont{P.}~\bibnamefont{{Papadopoulos}}},
  \bibnamefont{and}
  \bibinfo{author}{\bibfnamefont{J.}~\bibnamefont{{Winicour}}},
  \bibinfo{journal}{\cqg} \textbf{\bibinfo{volume}{14}}, \bibinfo{pages}{977}
  (\bibinfo{year}{1997}), \eprint{arXiv:gr-qc/9702002}.

\bibitem[{\citenamefont{{Jackson}}(1998)}]{Jackson}
\bibinfo{author}{\bibfnamefont{J.~D.} \bibnamefont{{Jackson}}},
  \emph{\bibinfo{title}{Classical Electrodynamics}}
  (\bibinfo{publisher}{Wiley-VCH}, \bibinfo{year}{1998}),
  \bibinfo{edition}{3rd} ed., ISBN \bibinfo{isbn}{047130932X}.

\bibitem[{\citenamefont{{Abramowitz} and {Stegun}}(1972)}]{AbramSteg}
\bibinfo{author}{\bibfnamefont{M.}~\bibnamefont{{Abramowitz}}}
  \bibnamefont{and} \bibinfo{author}{\bibfnamefont{I.~A.}
  \bibnamefont{{Stegun}}}, \emph{\bibinfo{title}{Handbook of Mathematical
  Functions}} (\bibinfo{publisher}{New York: Dover}, \bibinfo{year}{1972}).

\bibitem[{\citenamefont{{Friedrich}}(1988)}]{Friedrich1988}
\bibinfo{author}{\bibfnamefont{H.}~\bibnamefont{{Friedrich}}}, in
  \emph{\bibinfo{booktitle}{Highlights in Gravitation and Cosmology}}, edited
  by \bibinfo{editor}{\bibfnamefont{B.~R.} \bibnamefont{{Iyer}}},
  \bibinfo{editor}{\bibfnamefont{A.}~\bibnamefont{{Kembhavi}}},
  \bibinfo{editor}{\bibfnamefont{J.~V.} \bibnamefont{{Narlikar}}},
  \bibnamefont{and} \bibinfo{editor}{\bibfnamefont{C.~V.}
  \bibnamefont{{Vishveshwara}}} (\bibinfo{publisher}{Cambridge, UK, Cambridge
  University Press}, \bibinfo{year}{1988}).

\bibitem[{\citenamefont{{Geroch} and {Horowitz}}(1978)}]{GerochHorowitz}
\bibinfo{author}{\bibfnamefont{R.}~\bibnamefont{{Geroch}}} \bibnamefont{and}
  \bibinfo{author}{\bibfnamefont{G.~T.} \bibnamefont{{Horowitz}}},
  \bibinfo{journal}{\prl} \textbf{\bibinfo{volume}{40}}, \bibinfo{pages}{203}
  (\bibinfo{year}{1978}).

\bibitem[{\citenamefont{{Bishop} et~al.}(1997)\citenamefont{{Bishop},
  {G{\'o}mez}, {Lehner}, {Maharaj}, and {Winicour}}}]{News}
\bibinfo{author}{\bibfnamefont{N.~T.} \bibnamefont{{Bishop}}},
  \bibinfo{author}{\bibfnamefont{R.}~\bibnamefont{{G{\'o}mez}}},
  \bibinfo{author}{\bibfnamefont{L.}~\bibnamefont{{Lehner}}},
  \bibinfo{author}{\bibfnamefont{M.}~\bibnamefont{{Maharaj}}},
  \bibnamefont{and}
  \bibinfo{author}{\bibfnamefont{J.}~\bibnamefont{{Winicour}}},
  \bibinfo{journal}{\prd} \textbf{\bibinfo{volume}{56}}, \bibinfo{pages}{6298}
  (\bibinfo{year}{1997}), \eprint{arXiv:gr-qc/9708065}.

\bibitem[{\citenamefont{{Winicour}}(1988)}]{W1988}
\bibinfo{author}{\bibfnamefont{J.}~\bibnamefont{{Winicour}}}, in
  \emph{\bibinfo{booktitle}{Highlights in Gravitation and Cosmology}}, edited
  by \bibinfo{editor}{\bibfnamefont{B.~R.} \bibnamefont{{Iyer}}},
  \bibinfo{editor}{\bibfnamefont{A.}~\bibnamefont{{Kembhavi}}},
  \bibinfo{editor}{\bibfnamefont{J.~V.} \bibnamefont{{Narlikar}}},
  \bibnamefont{and} \bibinfo{editor}{\bibfnamefont{C.~V.}
  \bibnamefont{{Vishveshwara}}} (\bibinfo{publisher}{Cambridge, UK, Cambridge
  University Press}, \bibinfo{year}{1988}).

\bibitem[{\citenamefont{{Newman} and {Unti}}(1962)}]{NU1962}
\bibinfo{author}{\bibfnamefont{E.~T.} \bibnamefont{{Newman}}} \bibnamefont{and}
  \bibinfo{author}{\bibfnamefont{T.~W.~J.} \bibnamefont{{Unti}}},
  \bibinfo{journal}{\jmp} \textbf{\bibinfo{volume}{3}}, \bibinfo{pages}{891}
  (\bibinfo{year}{1962}).

\bibitem[{\citenamefont{{Janis} and {Newman}}(1965)}]{JN1965}
\bibinfo{author}{\bibfnamefont{A.~I.} \bibnamefont{{Janis}}} \bibnamefont{and}
  \bibinfo{author}{\bibfnamefont{E.~T.} \bibnamefont{{Newman}}},
  \bibinfo{journal}{\jmp} \textbf{\bibinfo{volume}{6}}, \bibinfo{pages}{902}
  (\bibinfo{year}{1965}).

\bibitem[{\citenamefont{{K\"uhnel}}(2006)}]{Kuehnel2003}
\bibinfo{author}{\bibfnamefont{W.}~\bibnamefont{{K\"uhnel}}},
  \emph{\bibinfo{title}{Differential Geometry, Curves Surfaces Manifolds}}
  (\bibinfo{publisher}{American Mathematical Society}, \bibinfo{year}{2006}),
  \bibinfo{edition}{2nd} ed., ISBN \bibinfo{isbn}{0821839888}.

\bibitem[{\citenamefont{{Winicour}}(2013)}]{W2013}
\bibinfo{author}{\bibfnamefont{J.}~\bibnamefont{{Winicour}}},
  \bibinfo{journal}{private communication}  (\bibinfo{year}{2013}).

\end{thebibliography}
\end{document}